\documentclass[10pt,journal,doublecolumn]{IEEEtran}

\usepackage{threeparttable,placeins,makecell,stfloats,cite}
\usepackage{threeparttable,arydshln}
\usepackage{amsmath,amssymb,bm}
\usepackage[dvips]{graphicx}
\usepackage[usenames]{color}
\usepackage{amsfonts}
\usepackage{latexsym}
\usepackage{multirow}
\usepackage{caption}
\usepackage{rotating}
\usepackage{float}
\usepackage[format=hang]{subfig}
\usepackage{url}
\usepackage{cite}

\usepackage{booktabs}
\usepackage{amsmath}
\usepackage{algorithm,algorithmic}
\usepackage{tikz}
\usepackage{nicematrix}
\usetikzlibrary{arrows.meta,calc,positioning,quotes,fit}
\usepackage{stfloats}

\usepackage{arydshln}
\usepackage[none]{hyphenat}

\usepackage{tikz}
\usetikzlibrary{matrix,fit,calc}

\begin{document}
	\title{Spatially Coupled Sparse Code Multiple Access (SC-SCMA): A Spectral Graph Approach}
	\author{Yiming~Gui, Zilong~Liu, Qu~Luo,~and~Pei~Xiao
		\thanks{
    Y. Gui is with the College of Electronics and Information Engineering, Shenzhen University, Shenzhen 518060, China (e-mail: ymgui\_18@163.com).
    Z. Liu is with the School of Computer Science and Electronics Engineering, University of Essex, UK (e-mail: zilong.liu@essex.ac.uk). 
    Qu  Luo  and  Pei  Xiao  are  with the Institute for Communications, University of Surrey, UK (email: \{q.u.luo,   p.xiao\}@surrey.ac.uk).
  }
	}
	\maketitle
	
\begin{abstract}
This paper presents a spatially coupled sparse code multiple access (SC-SCMA) framework to overcome the performance and scalability limitations of conventional SCMA systems. 
By analyzing the pairwise error probability associated to multi-user error patterns, we show that spatial coupling projects the superimposed SCMA codewords into a higher-dimensional effective signal space, leading to a strictly improved minimum Euclidean distance (MED) compared with conventional SCMA, while simultaneously enhancing the coding gain through global message propagation and the diversity gain through inter-block resource spreading. Such a distance gain is shown to be governed by the effective access dimensionality (EAD) induced by the coupled factor graph. With the aid of spectral graph theory, we establish a direct relationship between the spectral gap of the factor graph and a lower bound on the EAD, providing a computable structural metric that guarantees MED improvement under various error patterns. Building upon these theoretical insights, we introduce a low-complexity structure-aware codebook design approach, including a spectral-gap-oriented construction of spatially coupled factor matrices and a localized codebook optimization strategy that exploits the dominant error-inducing local user group. Simulation results validate the analysis and demonstrate that the proposed SC-SCMA consistently outperforms conventional SCMA in overloaded massive access channels.
\end{abstract}


\section{Introduction}
\IEEEPARstart{T}{he} widespread proliferation of machine-type communications (MTC), characterized by tens and billions of interconnected devices, sensors, robots, and vehicles, has necessitated non-orthogonal multiple access (NOMA) as an enabling technique for massive connectivity in next-generation wireless networks \cite{liuNonorthogonal2017,liuEvolutionNOMANext2022}.  Among many others, sparse code multiple access (SCMA) has emerged as a promising code-domain NOMA (CD-NOMA) \cite{liuSparseDenseComparative2021,yuSparseCodeMultiple2021} scheme. Unlike the conventional code-division multiple access (CDMA) approach, in which each user is allocated a dense spreading sequence, SCMA is built upon carefully designed sparse codebooks with favorable minimum-distance properties
 \cite{nikopourSparseCodeMultiple2013,taherzadehSCMACodebookDesign2014}. In SCMA, each sparse codebook corresponds to a multidimensional constellation and a sparse codeword is selected by a specific encoder (associated to certain user) according to the instantaneous input data bits. With the aid of message passing algorithm (MPA), the receiver is able to exploit the codebook sparsity as well as the constellation shaping gain to achieve bit error rate (BER) performance close to that of
 a maximum-likelihood receiver \cite{chaturvediTutorialDecodingTechniques2022}. 
This improves spectral efficiency and yields a distinct BER gain over its predecessor, called low-density spreading CDMA (LDS-CDMA), which relies on sparse spreading without explicit optimization of the multidimensional constellation \cite{hoshyarNovelLowDensitySignature2008}. 

Despite its promising performance, the effectiveness of SCMA systems critically depends on the quality of the underlying multidimensional sparse codebooks \cite{wenDesigningEnhancedMultidimensional2022,luoEnhancingSignalSpace2024}. Existing codebook design works mostly focus on optimizing constellation structures to maximize the minimum Euclidean/product distance (MED) \cite{yuDesignAnalysisSCMA2018,huangDownlinkSCMACodebook2022,caiMultiDimensionalSCMACodebook2016,liDesignPowerImbalancedSCMA2022}, improve the mutual information and constellation shaping gain \cite{sharmaSCMACodebookBased2018,dongEfficientSCMACodebook2018,maBitInterleavedCodedSCMA}, or enhance the codebook structural properties for low-complexity decoding \cite{bayestehLowComplexityTechniques2015,hanEnablingHighOrder2018,luoDesignLowProjectionSCMA2023}. While these methods provide improved error performance, they are effective only for small-scale SCMA systems. 

For a scalable deployment of SCMA, a common approach is to divide all the resource elements (i.e., subcarriers in a multi-carrier system) into multiple frequency blocks, each consisting of a small fixed number of resource elements, and then perform separate SCMA transmissions via orthogonal frequency division multiplexing (OFDM) by reusing the same set of small-scale codebooks. However, locally optimized small-sized codebooks may not yield further performance gains in large systems. This is because the isolated transmission scheme fails to facilitate a global interaction across different SCMA blocks, thereby limiting its capability to exploit the full channel diversity. Another approach is to directly design large-sized SCMA codebooks with favorable distance characteristics, yet this could result in prohibitively high complexity for both codebook optimization and MPA decoding and hence it may not be feasible in practice.   


To address the aforementioned challenges, we revisit the design philosophy of low-density parity-check (LDPC) codes. First, it is noted that the MPA based SCMA decoding performs iterative inference over a factor graph, sharing strong similarities with belief propagation decoding in LDPC theory, where the convergence behavior is governed by the global message propagation dynamics. Second, it is known that spatial coupling of multiple component factor graphs leads to significantly improved decoding thresholds and overall performance \cite{kudekarEffectSpatialCoupling2010,kudekarThresholdSaturationBMS2010}. In particular, spatially coupled LDPC codes connect multiple identical or similar component codes in a structured and overlapping manner, forming a coupled system that inherits and amplifies the advantages of its constituent parts \cite{schmalenCombiningSpatiallyCoupled,mitchellSpatiallyCoupledLDPC2015}. A key theoretical breakthrough, known as the threshold saturation phenomenon, demonstrates that such coupled systems can approach channel capacity under belief propagation decoding \cite{kudekarThresholdSaturationSpatial2011}. These observations suggest that introducing spatial coupling across multiple SCMA blocks may be an effective way to improve global message propagation and support scalable system construction. 

Inspired by the above observations, we propose a novel spatially coupled SCMA system (SC-SCMA) in this work. Instead of directly designing a large SCMA codebook, we propose to construct a spatially coupled structure based on a set of well-designed small constituent codebooks. By introducing structured spatial coupling across multiple constituent blocks, the resulting SC-SCMA system progressively approaches the behavior of a large optimized codebook while maintaining manageable design and decoding complexity. The main contributions of this work are summarized as follows: 

\begin{itemize}

\item 
We develop a unified pairwise error probability (PEP) based framework to analyze SC-SCMA under different multi-user error patterns. It is revealed that spatial coupling projects the superimposed SCMA codewords into a higher-dimensional signal space and thereby fundamentally reshapes the geometry of multi-user superposition constellation in SCMA systems, leading to a strict improvement of MED compared with conventional SCMA. We show that such a distance gain is governed by the effective access dimensionality (EAD) induced by the coupled factor graph, thereby establishing a direct link between signal-space dimensionality and error performance.
\item 
By leveraging spectral graph theory, we establish an analytical relationship between the spectral gap of the SCMA factor graph and a lower bound on the EAD. This finding provides a computable structural metric that guarantees the dimensionality expansion of the signal space and the associated MED improvement under multi-user error patterns and hence offers a rigorous graph-theoretic explanation of how spatial coupling enhances distance properties.
\item 
Building upon the aforementioned theoretical insights, we reformulate the spatially coupled factor matrix (SCFM) design as a spectral-gap-oriented optimization using a coupling assignment matrix. To facilitate efficient evaluation in large-scale systems, we introduce the concept of windowed SCFM (WSCFM)  which is shown to be sufficient for accurately capturing the spectral properties that govern global EAD behavior. Furthermore, by identifying the dominant error-inducing local user group (LUG) that determines the MED, the codebook optimization is reduced from a global high-dimensional design over all users to a LUG-based optimization, achieving substantial complexity reductions of 99.61\% and 99.99\% under $\varpi=150\%$ and $\varpi=200\%$, respectively, while preserving the favorable Euclidean distance structure of SC-SCMA. 

\end{itemize}

\subsection{Organization}
The remainder of this paper is organized as follows. Section II introduces the preliminaries of SCMA systems, including the conventional SCMA model, OFDM-based SCMA architecture, and the spectral-gap metric used to characterize graph connectivity. Section III presents the proposed SC-SCMA framework, including the construction of the SCFM, the corresponding transmission model, and the MPA-based detection algorithm. Section IV develops the theoretical foundation of SC-SCMA, where the relationships among PEP, EAD, MED, and spectral gap are established. Section V proposes a low-complexity design framework, including spectral-gap-oriented SCFM construction and localized codebook optimization based on the dominant error-inducing LUG. Section VI provides numerical results to validate the theoretical analysis and demonstrate the performance gains of the proposed SC-SCMA system. Finally, Section VII concludes this paper.

\subsection{Notations}
The $n$-dimensional complex and binary vector spaces are denoted as $\mathbb{C}^n$ and $\mathbb{B}^n$, respectively. Similarly, $\mathbb{C}^{K\times J}$ and $\mathbb{B}^{K\times J}$ denote the $(K\times J)$-dimensional complex and binary matrix spaces, respectively. $\mathbf{x}_v$ denotes the $v$-th column of matrix $\mathbf{X}$. $x_{k,v}$ or $\mathbf{X}[k,v]$ represent the element located at the $k$-th row and $v$-th column of $\mathbf{X}$. $\text{diag}(\mathbf{x})$ returns a diagonal matrix whose diagonal entries are given by $\mathbf{x}$. $(\cdot)^T$ denotes the transpose operation.  

\section{Preliminaries}
\subsection{SCMA}
Consider a $K \times J$ SCMA system, where $J$ users simultaneously communicate over $K$ resource nodes (RNs). The overloading factor of the system is defined as  $\varpi = J/K>1$, which indicates that the number of active users exceeds the number of available orthogonal resources. Each user is assigned a sparse codebook (denoted by $\mathcal{X}$) and there are $J$ codebooks in total, that is
\begin{equation}
   \mathcal{X} =\{\mathbf{X}_1,\mathbf{X}_2,\cdots,\mathbf{X}_J\},
\end{equation}
where $\mathbf{X}_j=[\mathbf{x}_{j,1},\mathbf{x}_{j,2},\cdots,\mathbf{x}_{j,M}]$ denotes the codebook set for user $j$, $M$ is the codebook size, and $\mathbf{x}_{j,m}=[x_{1,m},x_{2,m},\cdots,x_{K,m}]$ is the $m$-th $k$-dimensional codeword. The average codeword power of a codebook is assumed to be unit, i.e., $\frac{1}{M}=\sum_{m=1}^{M}\| \mathbf{x}_{j,m} \|^2=1$.

\begin{figure*}
\centering
\includegraphics[width=5in]{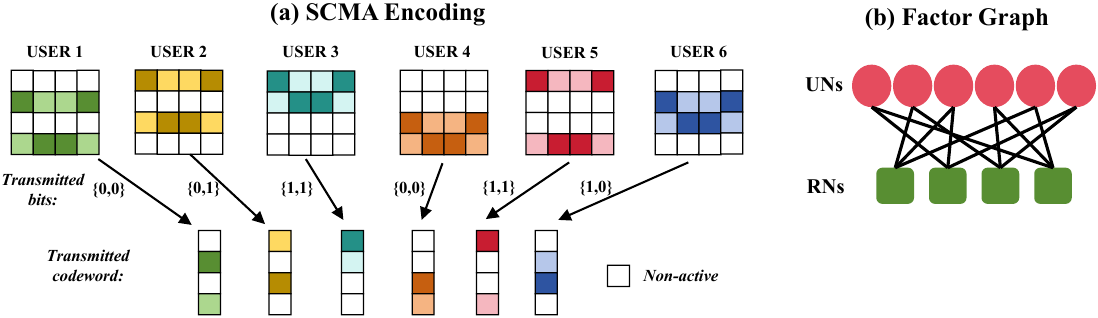} \\
\caption{Illustration of SCMA Codeword Mapping and Factor Graph Representation.}
\label{fig:SCMA}
\end{figure*}

The basic principle of SCMA encoding is illustrated in Fig. \ref{fig:SCMA}(a). At the transmitter, each user encodes $\log_2 M$ bits into a $K$-dimensional sparse codeword. Denoted $\mathbf{b}_j=[b_{j,1},b_{j,2},\cdots,b_{j,\log_2M}]$ by the input binary message of $j$-th user. The mapping process can be written as
\begin{equation}
    g_j: \mathbb{B}^{\log_2 M \times 1} \rightarrow \mathbf{X}_j\in \mathbb{C}^{K\times M},\ \text{i.e.},\ \mathbf{x}_j=g_j(\mathbf{b}_j).
\end{equation}

The sparse resource allocation of SCMA can be captured by a binary factor matrix, that is
\begin{equation}\label{eq:prototype_fm}
 \mathbf{F}_{4\times 6}
=
\begin{bmatrix}
0 & 1 & 1 & 0 & 1 & 0\\
1 & 0 & 1 & 0 & 0 & 1\\
0 & 1 & 0 & 1 & 0 & 1\\
1 & 0 & 0 & 1 & 1 & 0
\end{bmatrix},
\end{equation}
where $f_{k,j}=1$ indicates that the $j$-th user is connected to the $k$-th RN, and $f_{k,j}=0$ otherwise. The corresponding factor graph representation is shown in Fig. \ref{fig:SCMA}(b). For a user node (UN) $j$, its neighborhood is defined as
\begin{equation}
\mathcal{N}_{u}(j)
=
\left\{
k \,\middle|\, f_{k,j}=1
\right\},
\end{equation}
which represents the set of RNs occupied by UN $j$. Similarly, for an RN $k$, its neighborhood is defined as
\begin{equation}
\mathcal{N}_{r}(k)
=
\left\{
j \,\middle|\, f_{k,j}=1
\right\},
\end{equation}
which represents the set of UNs sharing RN $k$. For regular SCMA, each UN is connected to exactly $d_v$ RNs and each RN is connected to exactly $d_f$ UNs. Therefore,
\begin{equation}
|\mathcal{N}_{u}(j)| = d_v,
\quad
|\mathcal{N}_{r}(k)| = d_f.
\end{equation}

By counting the total number of edges in the factor graph from both the UN and RN sides, we obtain
\begin{equation}
Jd_v = Kd_f.
\end{equation}

In the following, the factor matrix in (\ref{eq:prototype_fm}) for a basic SCMA system is referred to as the prototype factor matrix, denoted by $\mathbf{F}_{p}$, which fully characterizes the local connectivity pattern of the underlying SCMA system and provides the fundamental building block for constructing the global factor matrix introduced in the next subsection.

\subsection{OFDM-SCMA}
\begin{figure}
\centering
\includegraphics[width=3.3in]{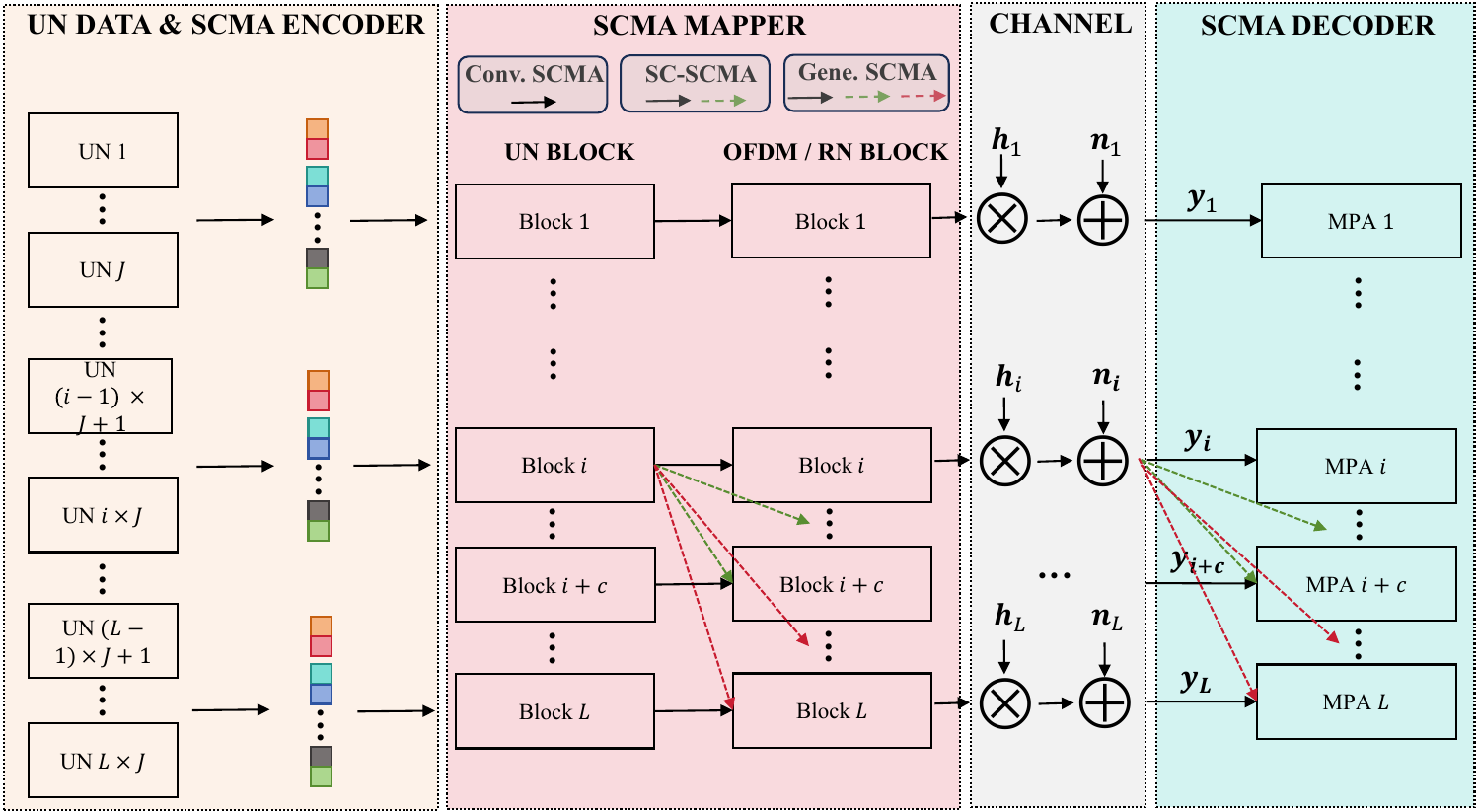} \\
\caption{OFDM-SCMA architecture featuring various SCMA mapping strategies and their corresponding coupling-based MPA decoding.}
\label{fig:system_model}
\end{figure}

Consider an OFDM-SCMA system, where the basic SCMA blocks are mapped onto the available OFDM subcarriers as illustrated in Fig.~\ref{fig:system_model}. Let the total number of OFDM subcarriers be $N = LK$, which are partitioned into $L$ SCMA resource blocks, each containing $K$ RNs and supporting $J$ users according to the prototype factor matrix $\mathbf{F}_{p}$.

For the $l$-th resource block, where $l\in{1,\ldots,L}$, the transmitted
SCMA codeword set is given by

\begin{equation}
\left[
(\mathbf{x}^{(l)}_{1})^T,
(\mathbf{x}^{(l)}_{2})^T,
\ldots,
(\mathbf{x}^{(l)}_{J})^T
\right]^T,
\end{equation}
where $\mathbf{x}^{(l)}_{j}\in\mathbb{C}^{K\times1}$ denotes the codeword
transmitted by user $j$ over the $l$-th resource block.

Since the same connectivity pattern is employed across all resource blocks,
the overall OFDM-SCMA factor graph consists of $L$ independent replicas of
the prototype SCMA graph. Accordingly, the corresponding extended factor
matrix can be expressed as
\begin{equation}
\mathbf{F}_{\mathrm{OFDM}} = \mathbf{I}_{L}\otimes\mathbf{F}_p=
\begin{bmatrix}
\mathbf{F}_p & \mathbf{0} & \cdots & \mathbf{0} \\
\mathbf{0} & \mathbf{F}_p & \cdots & \mathbf{0} \\
\vdots & \vdots & \ddots & \vdots \\
\mathbf{0} & \mathbf{0} & \cdots & \mathbf{F}_p
\end{bmatrix},
\end{equation}
where $\mathbf{F}_{\mathrm{OFDM}}$ denotes the extended factor matrix of OFDM-SCMA. $\mathbf{I}_{L}$ is the $L\times L$ identity matrix and $\otimes$
denotes the Kronecker product.

\subsection{Spectral Gap} 
The extended factor matrix $\mathbf{F}_{\mathrm{OFDM}}$ of a conventional OFDM-SCMA system exhibits a block-diagonal structure, where each diagonal block corresponds to an independent SCMA resource block. As a result, no edges exist between different transmission blocks, and the overall graph is composed of multiple disconnected components.

To facilitate the analysis of the overall graph structure, we associate $\mathbf{F}$ with the bipartite adjacency matrix
\begin{equation}
\mathbf{A}=
\begin{bmatrix}
\mathbf{0} & \mathbf{F}\\
\mathbf{F}^{T} & \mathbf{0}
\end{bmatrix}.
\end{equation}

The adjacency matrix $\mathbf{A}$ provides an equivalent graph representation of the connectivity induced by $\mathbf{F}$. Furthermore, its eigenvalue spectrum enables the application of spectral graph theory to characterize global connectivity properties of the resulting OFDM-SCMA graph. Let ${\mu_1,\mu_2,\ldots,\mu_n}$ denote the eigenvalues of $\mathbf{A}$ sorted in descending order. A key metric of interest is the spectral gap, defined as the normalized difference between the largest and the second-largest eigenvalues
\begin{equation}
    \Delta = \frac{\mu_1 - \mu_2}{\mu_1} \geq 0.
\end{equation}

The spectral gap characterizes the separation between the dominant connectivity mode and the remaining structural modes of the graph. In
general, a larger spectral gap indicates stronger global connectivity, whereas a small spectral gap implies the presence of weakly connected or
disconnected graph components. For the conventional OFDM-SCMA system, the global factor graph consists of $L$ independent SCMA blocks. Accordingly, the corresponding adjacency matrix can be written as a block-diagonal matrix
\begin{equation}
\mathbf{A}_{\mathrm{OFDM}} =
\begin{bmatrix}
\mathbf{A}_p & \mathbf{0} & \cdots & \mathbf{0} \\
\mathbf{0} & \mathbf{A}_p & \cdots & \mathbf{0} \\
\vdots & \vdots & \ddots & \vdots \\
\mathbf{0} & \mathbf{0} & \cdots & \mathbf{A}_p
\end{bmatrix},
\end{equation}
where $\mathbf{A}_p$ is the adjacency matrix of prototype matrix $\mathbf{F}_p$.

A standard result from linear algebra is that the spectrum of a block-diagonal matrix is the union of the spectra of its diagonal blocks. Therefore,
\begin{equation}
    \mathrm{spec}(\mathbf{A}_{\mathrm{OFDM}})=\bigcup_{\ell=1}^L \mathrm{spec}(\mathbf{A}_{p}).
\end{equation}

Hence the largest eigenvalue $\mu_{\max}$ of $\mathbf{A}_p$ appears $L$ times in the spectrum of $\mathrm{spec}(\mathbf{A}_{\mathrm{OFDM}})$. Consequently, 

\begin{equation}
\mu_1=\mu_2=\cdots=\mu_L=\mu_{\max},
\end{equation}
and therefore $\Delta=0$. The vanishing spectral gap reveals that the conventional OFDM-SCMA graph is disconnected, thereby limiting information propagation across transmission blocks. To overcome this limitation, we introduce spatial coupling, whereby adjacent SCMA blocks are connected through structured inter-block edges. According to a classical result in spectral graph theory, the largest eigenvalue of the adjacency matrix is simple for a connected graph, yielding
\begin{equation}
\mu_1 > \mu_2,
\end{equation}
and hence
\begin{equation}
\Delta>0.
\end{equation}

The resulting increase in spectral gap indicates improved global connectivity. The corresponding spatially coupled construction is presented in the next section.

\section{Proposed SC-SCMA}\label{section:3}
In this section, we present the proposed SC-SCMA framework based on spatial coupling. First, the construction of the SCFM and its windowed representation are introduced, providing a locally structured yet globally connected graph architecture. Then, the downlink SC-SCMA transmission model is established based on the proposed coupling mechanism. Finally, the corresponding MPA-based detection scheme is developed.

\subsection{SCFM and Windowed SCFM}
As depicted in Fig. \ref{fig:system_model}, we introduce spatial coupling to improve the global connectivity of the conventional OFDM-SCMA structure. When each block is connected to its subsequent $c$ blocks, the resulting structure is referred to as $c$-level spatial coupling. The maximum coupling level $c$ is constrained by the column weight $d_v$ of the prototype matrix, which equals the number of RNs occupied by each user. In this work, we consider $d_v=2$ and therefore focus on coupling levels up to $c=2$. As a result, the extended factor matrix evolves from a block-diagonal structure to a banded structure, enabling interactions among RNs across neighboring blocks and forming a globally connected graph.

To realize spatial coupling while preserving the sparsity of the factor matrix, the prototype matrix $\mathbf{F}_p$ is decomposed into $c+1$ binary component matrices $\mathbf{W}_i\in \mathbb{B}^{K\times J}$ according to
\begin{equation}
\mathbf{F}_{p}=\sum_{i=0}^{c}\mathbf{W}_i.
\end{equation}

Each component matrix $\mathbf{W}_i$ represents the $i$-th subset of user-resource connections in $\mathbf{F}_p$. Specifically, the nonzero entries of $\mathbf{W}_{i}$ correspond to edges that are connected to the $i$-th neighboring block in the spatially coupled structure. Since every nonzero entry of $\mathbf{F}_p$ appears exactly once among $\{\mathbf{W}_{i}\}_{i=0}^{c}$, the original sparsity pattern and node degrees are preserved.

Based on this decomposition, the SCFM is obtained by placing each component matrix $\mathbf{W}_{i}$ at the corresponding inter-block offset. The matrix $\mathbf{W}_{0}$ maintains intra-block connections, whereas $\mathbf{W}_{i}$ for $i\geq1$ introduces controlled inter-block connections. Consequently, the resulting SCFM exhibits a banded block structure in which each block row interacts only with a limited number of neighboring block columns, which can be expressed as
\begin{equation}\label{eq:scfm}
\mathbf{F} =
\begin{tikzpicture}[baseline=(M.center)]
\matrix (M) [matrix of math nodes,left delimiter={[},right delimiter={]}]
{
\ddots & \ddots & \ddots & \ddots & \ddots & \ddots & \ddots \\
\ddots & \mathbf{W}_0 & \ddots & \ddots & \ddots & \ddots & \ddots \\
\ddots & \vdots & \ddots & \ddots & \ddots & \ddots & \ddots \\
\ddots & \mathbf{W}_c & \cdots & \mathbf{W}_0 & \ddots & \ddots & \ddots \\
\ddots & \ddots & \ddots & \vdots & \ddots & \ddots & \ddots \\
\ddots & \ddots & \ddots & \mathbf{W}_c & \cdots & \mathbf{W}_0 & \ddots \\
\ddots & \ddots & \ddots & \ddots & \ddots & \vdots & \ddots \\
\ddots & \ddots & \ddots & \ddots & \ddots & \mathbf{W}_c & \ddots \\
\ddots & \ddots & \ddots & \ddots & \ddots & \ddots & \ddots \\
};

\node[draw=red,thick,rounded corners,fit=(M-2-2)(M-8-6),
      inner sep=2pt] (box) {};

\node[red,above right] at (box.north east) {$\mathbf{F}_{w}$};

\end{tikzpicture},
\end{equation}
where the red-boxed region in (\ref{eq:scfm}) defines the windowed SCFM, denoted by $\mathbf{F}_{w}$. The matrix $\mathbf{F}_{w}$ is obtained by restricting the SCFM to the finite coupling window associated with block $\ell$, while excluding interactions outside this region.

\begin{figure*}
\centering
\includegraphics[width=6in]{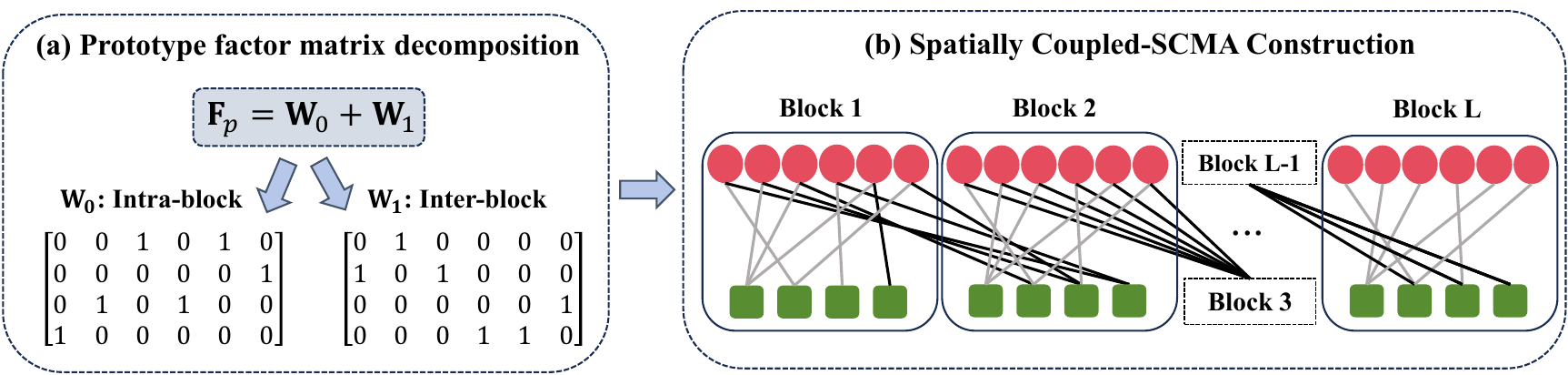} \\
\caption{Construction of Spatially Coupled-SCMA via prototype matrix decomposition.}
\label{fig:scfm_construction}
\end{figure*}
To illustrate the construction of $\mathbf{F}_{w}$, consider the representative case of $c=1$. As can be shown in Fig. \ref{fig:scfm_construction}(a), the prototype matrix is decomposed into two binary component matrices corresponding to intra-block and inter-block connections, respectively

\begin{equation}\label{eq:BF1}
\mathbf{W}_{0} = \begin{bmatrix}
    0 & 0 & 1 & 0 & 1 & 0 \\
    0 & 0 & 0 & 0 & 0 & 1 \\
    0 & 1 & 0 & 1 & 0 & 0 \\
    1 & 0 & 0 & 0 & 0 & 0 \\
\end{bmatrix},
\end{equation}

\begin{equation}\label{eq:BF2}
\mathbf{W}_{1} = \begin{bmatrix}
    0 & 1 & 0 & 0 & 0 & 0 \\
    1 & 0 & 1 & 0 & 0 & 0 \\
    0 & 0 & 0 & 0 & 0 & 1 \\
    0 & 0 & 0 & 1 & 1 & 0 \\
\end{bmatrix}.
\end{equation}

Let $\{\mathcal{R}_\ell\}_{\ell=1}^L$ and $\{\mathcal{U}_\ell\}_{\ell=1}^{L}$ denote a partition of the RNs and UNs into $L$ disjoint blocks, where $|\mathcal{R}_\ell|=K$ and $|\mathcal{U}_\ell|=J$. For an RN block $\mathcal{R}_\ell$, we define its block-level neighborhood as 
\begin{equation}
\mathcal{N}_r(\mathcal{R}_\ell)=\bigcup_{k\in\mathcal{R}_\ell}
\mathcal{N}_{r}(k),
\end{equation}
which represents the set of UNs connected to at least one RN in block $\mathcal{R}_\ell$.

For the conventional OFDM-SCMA system, each RN block interacts exclusively with the corresponding UN block, yielding
\begin{equation}
\mathcal{N}_r(\mathcal{R}_\ell)=\mathcal{U}_\ell,
\quad
\ell=1,\ldots,L.
\end{equation}

In the spatially coupled system, however, each RN block may interact with multiple neighboring UN blocks. To characterize the local connectivity induced by windowed SCFM, we define a windowed RN block centered at $\mathcal R_\ell$ as
\begin{equation}
\mathcal{R}_\ell^w=\bigcup_{i=-c}^{i=c} \mathcal{R}_{\ell+i}.
\end{equation}

The corresponding set of neighboring UNs is then given by
\begin{equation}
\mathcal{N}_r(\mathcal{R}_\ell^w)=\bigcup_{i=-c}^{i=2c} \mathcal{N}_r(\mathcal{R}_{\ell+i}).
\end{equation}

Accordingly, the windowed factor matrix $\mathbf F_w$ is defined as the submatrix of the SCFM induced by the RN blocks contained in
$\mathcal R_\ell^w$ and their neighboring UN blocks $\mathcal N_u(\mathcal R_\ell^w)$, which can be expressed as 
\begin{equation}
\mathbf{F}_{w}
\triangleq
\mathbf{F}\bigl[\mathcal{R}_\ell^w,\mathcal{N}_r(\mathcal{R}_\ell^w)\bigr].
\end{equation}

The windowed matrix $\mathbf{F}_{w}$ serves as the fundamental structural unit of the spatially coupled system, since it completely characterizes the local connectivity pattern induced by the coupling mechanism. By sliding this window along the coupling chain, the entire SCFM can be viewed as a sequence of overlapping local structures connected through shared nodes.

The corresponding factor graph representation of the windowed SCMA system is illustrated in Fig. \ref{fig:scfm_construction}(b). As shown in the figure, each transmission block is associated with a local coupling window $\mathbf{F}_{w}$ that overlaps with adjacent blocks according to the coupling memory $c$. These overlapping windows introduce structured inter-block connections, transforming the originally disconnected block-diagonal graph into a connected graph. Consequently, information can propagate across the coupling chain through successive local interactions, resulting in an increased spectral gap and enhanced global connectivity.


\subsection{Downlink SC-SCMA}
The proposed SC-SCMA system consists of $L$ coupled transmission blocks. Let $\mathbf{x}_{j}^{(\ell)}\in\mathbb{C}^{K\times1}$ denote the sparse
SCMA codeword transmitted by user $j$ in block $\ell$, where the nonzero positions are determined by the windowed SCFM.

For each block $\ell$, the superimposed transmit vector is given by

\begin{equation}
\mathbf{x}^{(\ell)}
=
\sum_{j\in \mathcal{N}_r(\mathcal{R}_l)}
\mathbf{x}_{j}^{(\ell)},
\end{equation}
where $\mathcal{N}_r(\mathcal{R}_l)$ denotes the set of active user nodes associated with block $\ell$.

By stacking all transmission blocks, the overall transmitted signal can be
expressed as

\begin{equation}
\mathbf{x}=\left[(\mathbf{x}^{(1)})^{T},(\mathbf{x}^{(2)})^{T},\cdots,(\mathbf{x}^{(L)})^{T}\right]^T.
\end{equation}

Let $\mathbf{h}$ denote the equivalent channel vector. The received signal
can be written as

\begin{equation}
\mathbf{y}
=
\mathrm{diag}(\mathbf{h})\cdot \mathbf{x}
+
\mathbf{n},
\end{equation}
where $\mathbf{n}\sim\mathcal{CN}(\mathbf{0},N_0\mathbf{I})$ denotes the
additive white Gaussian noise vector.

\subsection{SC-SCMA Detection} 	
SC-SCMA preserves the sparsity and node degrees of the original factor graph. Consequently, the local computational complexity of message passing remains unchanged, and the MPA can be directly applied for multiuser detection. Unlike conventional OFDM-SCMA, however, the spatial coupling structure enables messages to propagate across neighboring transmission blocks through the inter-block connections introduced by the windowed SCFM, thereby facilitating information exchange over a larger graph.

Similar to many previous works, perfect channel state information (CSI) is assumed to be available at the receiver. The MPA performs iterative message updates between RNs and UNs until either convergence is achieved or a predefined maximum number of iterations is reached. During initialization, all codewords are assumed to be equiprobable, and the initial message from RN $k$ to UN $j$ is given by
\begin{equation}
\eta_{k \rightarrow j}^{(0)}(m)=\frac{1}{M},
\quad
m=1,\ldots,M.
\end{equation}

The subsequent message updates are performed according to the standard MPA rules described below.
\begin{equation}
\begin{split}
    \eta_{k\to j}^{(t)}(\mathbf{x}_j)\!&=\!\sum_{\sim\mathbf{x}_j}\frac{1}{\sqrt{\pi N_0}}\exp \\
    &\left\{- \frac{\left(y_{k}-h_{k}\sum_{v\in \mathcal{N}_r(k)}x_{k,v}\right)^2}{N_0}\right\} \\
    & \times \prod\limits_{r\in \mathcal{N}_r(k)\setminus\{v\}}\eta_{r\to k}^{(t-1)}(\mathbf{x}_r), \\
\end{split}
\end{equation}
where $\eta_{k\to j}^{(t)}$ denotes the belief message passing from RN $k$ to UN $j$ in iteration $t$. $\mathcal{N}_r(k)\setminus \left\{v\right\}$ denotes all the UNs in RN $k$ except UN $v$.

The message update rule at the UNs is given by 
\begin{equation}
    \eta_{j\to k}^{(t)}(x_{j})=\prod\limits_{r\in \mathcal{N}_u(j)\setminus\left\{k\right\}} \eta_{r\to j}^{(t-1)},
\end{equation}
where $\eta_{j\to k}^{(t)}$ denotes the extrinsic belief message transmitted from UN $j$ to RN $k$ during the $t$-th iteration, and $\mathcal{N}_u(j)\setminus \left\{k\right\}$ indicates the set of all RNs connected to UN $j$ excluding RN $k$.

\section{Theoretical Foundation of SC-SCMA}\label{subsection:3}
\begin{figure}
	\centering
	\includegraphics[width=3.5in]{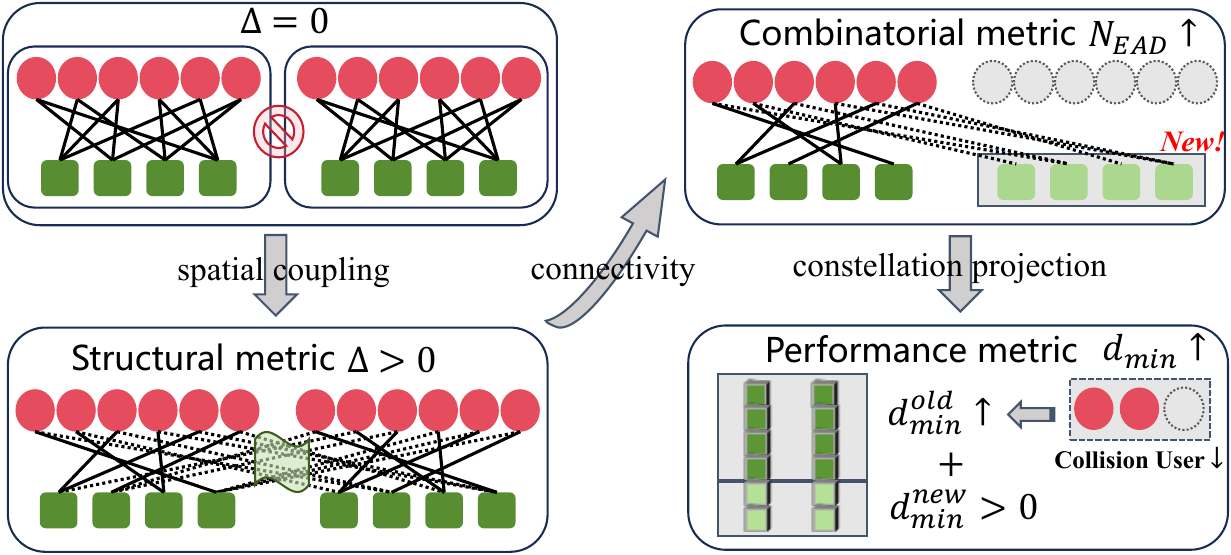} \\
	\caption{Structural-to-Performance Mapping in SC-SCMA.}
	\label{fig:mech_analysis}
\end{figure}

In this section, as depicted in Fig. \ref{fig:mech_analysis}, we analyze the error performance of SC-SCMA in terms of PEP and its structural dependence on the EAD. We show that the minimum Euclidean distance increases linearly with the EAD, leading to an exponential decay in the PEP. Furthermore, we establish a spectral graph-theoretic interpretation, demonstrating that the EAD is lower bounded by the spectral gap of the factor graph. This reveals that spatial coupling improves error performance by enlarging the spectral gap, increasing the EAD, and consequently reducing the PEP.

\subsection{PEP Analysis for SC-SCMA}\label{subsection:3-1}
First, consider a $V$-user error event associated with a user subset $\mathcal{V}$ selected from the total $J_{\mathrm{tot}}=LJ$ users. Let
\begin{equation}
\mathcal{N}_u(\mathcal{V})
=
\bigcup_{v\in\mathcal{V}}
\mathcal{N}_u(v),
\end{equation}
denote the set of RNs participating in the error event. Since the PEP depends on the signal differences accumulated over these RNs, the error performance is fundamentally governed by the size of $\mathcal{N}_u(\mathcal{V})$. Since each participating RN contributes one independent signal dimension, the effective signal-space dimensionality is equivalently measured by the cardinality of the participating RN set. Therefore, we define the EAD as
\begin{equation}
N_{\mathrm{EAD}}
=
\left|
\mathcal{N}_u(\mathcal{V})
\right|.
\end{equation}

For a maximum-likelihood receiver, the conditional PEP can be expressed as
\begin{equation}
\begin{split}
    \text{Pr}\{\mathcal{\mathbf{S}}_i \to &\mathcal{\mathbf{S}}_j \mid \mathbf{h}\}
    = \mathbb{E}_{\mathbf{h}}\left[ Q\left( \sqrt{\frac{\left\| \text{diag}(\mathbf{h})(\mathcal{\mathbf{s}}_i-\mathcal{\mathbf{s}}_j) \right\|_2^2}{2N_0}} \right) \right] \\
    &= \mathbb{E}_{\mathbf{h}}\left[ Q\left( \sqrt{\frac{\sum_{k\in \mathcal{N}_u(\mathcal{V})} h_k^2 \|s_{k,i}-s_{k,j}\|_2^2}{2N_0}} \right) \right] \\
    &\leq \mathbb{E}_{\mathbf{h}}\left[ \exp\left(-\frac{\sum_{k\in \mathcal{N}_u(\mathcal{V})} h_k^2 \|s_{k,i}-s_{k,j}\|_2^2}{4N_0} \right) \right],
\end{split}
\end{equation}
where $Q(\cdot)$ denotes the Gaussian $Q$-function, with $Q(x)=(2\pi)^{-1/2}\int_x^{+\infty}\exp(-t^2/2)\mathrm{d}t$, and $Q(x)\leq \exp(-x^2/2)$. 

Under an addictive white Gaussian noise (AWGN) channel\footnote{The PEP bound above assumes an AWGN channel, which is the appropriate model for interference-limited SCMA systems operating at high overloading factors. In this regime, the dominant impairment is multi-user interference rather than channel fading, and the error performance is well characterized by the Euclidean distance structure of the superimposed constellation. The AWGN analysis therefore captures the true performance bottleneck in the overloaded scenarios considered in this work.}, i.e., $h_k=1$ for all $k$, the PEP reduces to
\begin{equation}\label{eq:PEP-for-AWGN}
    \text{Pr}\{\mathcal{\mathbf{s}}_i \to \mathcal{\mathbf{s}}_j\}
    \leq \exp\left(-\frac{\sum_{k\in \mathcal{N}_u(\mathcal{V})} \|s_{k,i}-s_{k,j}\|_2^2}{4N_0} \right).
\end{equation}

\begin{figure}
	\centering
	\includegraphics[width=3.5in]{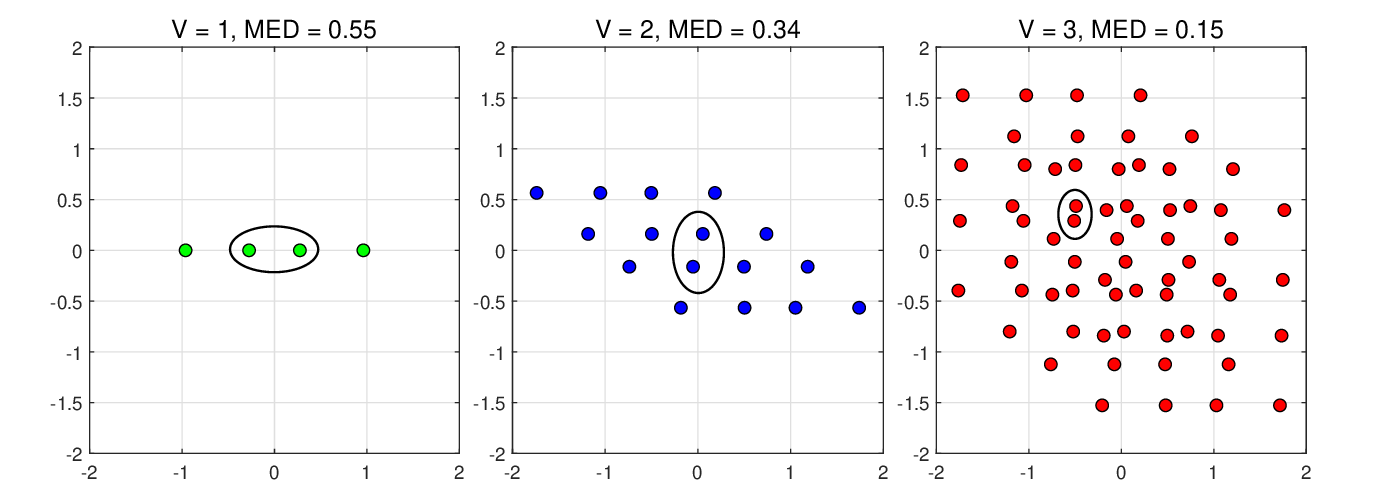} \\
	\caption{Superposed constellation of multiple colliding users on a single RN using Huawei codebook \cite{altera_scma_env1}.}
	\label{fig:med_evolution}
\end{figure}

Define
\begin{equation}
\tau_{\min}
=
\min_{k\in\mathcal{N}_u(\mathcal V)}
\;
\min_{i\neq j}
\|s_{k,i}-s_{k,j}\|_2^2,
\end{equation}
which represents the MED of the projected constellation on a single RN as show in Fig. \ref{fig:med_evolution}.

Accordingly, the MED associated with the $V$-user error event is
\begin{equation}
d_{\min}(\mathcal V)
=
\min_{i\neq j}
\sum_{k\in\mathcal{N}_u(\mathcal V)}
\|s_{k,i}-s_{k,j}\|_2^2.
\end{equation}

By construction, each participating RN contributes at least $\tau_{\min}$ to the accumulated distance, yielding
\begin{equation}
d_{\min}(\mathcal V)
\ge
N_{\mathrm{EAD}}\cdot
\,\tau_{\min}.
\label{eq:med_bound}
\end{equation}

Equation (\ref{eq:med_bound}) shows that the MED grows at least linearly with the EAD. Substituting this bound into (\ref{eq:PEP-for-AWGN})
reveals that increasing $N_{\mathrm{EAD}}$ leads to an exponential reduction in the PEP.

The gain of SC-SCMA can be understood as two complementary effects operating over different channel regimes. In AWGN and interference-limited channels, the enlarged EAD increases the accumulated Euclidean distance, directly boosting the coding gain as captured by the MED bound in (\ref{eq:med_bound}). In frequency-selective fading channels, the inter-block connectivity spreads each user's resource occupancy across a larger portion of the available bandwidth. Therefore, the coupled codewords experience more statistically independent fading realizations, increasing the effective frequency diversity order. These two effects are not independent: both are governed by the EAD, a larger EAD simultaneously means more distance dimensions (coding gain) and more independently faded resource contributions (diversity gain).


\subsection{Spectral Characterization of the EAD}\label{subsection:3-2}
In this subsection, we establish an explicit relationship between spatial coupling and the EAD from a spectral graph perspective. Specifically, we show that enlarging the spectral gap of the factor graph increases the minimum number of RNs participating in a multi-user error event.

Let $J_{\mathrm{tot}}$ denote the total number of UNs. Consider a $V$-user error event associated with a user subset $\mathcal{V} \subseteq \{1,\ldots,J_{\mathrm{tot}}\}$. Define the indicator vector
\begin{equation}
\mathbf{g}=[g_1,\ldots,g_{J_{\mathrm{tot}}}]^T\in\mathbb{B}^{J_{\mathrm{tot}}\times 1},
\end{equation}
where $g_j=1$ if $j\in\mathcal V$ and $g_j=0$ otherwise. 

The RN occupancy vector is defined as
\begin{equation}
\mathbf q = \mathbf F \mathbf g.
\end{equation}

The $k$-th entry of $\mathbf q$ can be written as
\begin{equation}
q_k
=
\sum_{j=1}^{J_{\mathrm{tot}}} f_{k,j} g_j
=
\left|
\mathcal N_r(k)\cap \mathcal V
\right|,
\end{equation}
which represents the number of users that occupy RN $k$.

By definition, $\mathbf{q}$ contains exactly $N_{\text{EAD}}$ nonzero entries, whose positions form the set $\mathbf{\Gamma}$ corresponding to the neighborhood of $\mathcal{V}$ in the bipartite graph. Moreover, the total number of active edges satisfies
\begin{equation}
\|\mathbf{q}\|_1 = \sum_{i \in \mathbf{\Gamma}} q_i = V d_v.
\end{equation}

Applying the Cauchy-Schwarz inequality over $\mathbf q$ yields
\begin{equation}\label{q_with_gamma}
\begin{split}
    \|\mathbf{q}\|_1
    &= N_{\text{EAD}} \cdot \frac{1}{N_{\text{EAD}}}\sum_{i\in \mathbf{\Gamma}} q_i \\
    &\leq N_{\text{EAD}} \sqrt{\sum_{i\in \mathbf{\Gamma}} q_i^2 \cdot \sum_{i\in \mathbf{\Gamma}} N_{\text{EAD}}^{-2}} \\
    &= \sqrt{N_{\text{EAD}} \|\mathbf{q}\|_2^2}.
\end{split}
\end{equation}

Consequently,
\begin{equation}\label{q_with_gamma_final}
    N_{\text{EAD}} \geq \frac{\|\mathbf{q}\|_1^2}{\|\mathbf{q}\|_2^2} = \frac{V^2 d_v^2}{\|\mathbf{q}\|_2^2},
\end{equation}
which shows that for a fixed number of active edges, a smaller value of $\|\mathbf q\|_2^2$ leads to a larger EAD.

To bound $\|\mathbf{q}\|_2^2$, consider the eigenvalue decomposition of $\mathbf F^T\mathbf F$. Let 
\begin{equation}
    \mathbf{g} = \sum_{i=1}^{J_{\text{tot}}} \beta_i \mathbf{e}_i
\end{equation}
where $\mathbf e_i$ is the eigenvector associated with eigenvalue $\lambda_i$ of $\mathbf F^T\mathbf F$, ordered as
\begin{equation}
\lambda_1
\ge
\lambda_2
\ge
\cdots
\ge
0,
\end{equation}
where $\mathbf{e}_i$ are the eigenvectors of $\mathbf{F}^T\mathbf{F}$. We have
\begin{equation}
\sum_{i=1}^{J_{\text{tot}}} \beta_i^2 = \|\mathbf{g}\|_2^2 = V, \quad
\beta_1 = \frac{V}{\sqrt{J_{\text{tot}}}},
\end{equation}

Then,
\begin{equation}\label{eq:q_square}
\begin{split}
    \|\mathbf{q}\|_2^2
    &= \mathbf{g}^T \mathbf{F}^T \mathbf{F} \mathbf{g}
    = \sum_{i=1}^{J_{\text{tot}}} \lambda_i \beta_i^2 \\
    &\leq \lambda_1 \beta_1^2 + \lambda_2 \sum_{i=2}^{J_{\text{tot}}} \beta_i^2 \\
    &= V\left( \lambda_1 \frac{V}{J_{\text{tot}}} + \lambda_2 \left(1 - \frac{V}{J_{\text{tot}}}\right) \right).
\end{split}
\end{equation}

Combining the (\ref{eq:q_square}) and (\ref{q_with_gamma_final}), we obtain
\begin{equation}\label{eq:ead_mid}
N_{\text{EAD}} \geq \frac{N}{1 + \frac{\lambda_2}{\lambda_1} \cdot \frac{J_{\text{tot}} - V}{V}}.
\end{equation}
where $\lambda_1 = d_f d_v$, which can refer to Appendix \ref{app:a}.

Next, we establish the connection between the eigenvalues of $\mathbf F^T\mathbf F$ and the spectral gap of the adjacency matrix.
Specifically, let
\begin{equation}\label{eq:B_FFT}
\mathbf{A} 
\begin{bmatrix} 
\mathbf{u} \\ 
\mathbf{v} 
\end{bmatrix}
= \mu 
\begin{bmatrix} 
\mathbf{u} \\ 
\mathbf{v} 
\end{bmatrix},
\end{equation}
where $\mathbf{u} \in \mathbb{R}^{KL}$ and $\mathbf{v} \in \mathbb{R}^{JL}$ denote the components of the eigenvector associated with the RNs  and UNs of the bipartite graph, respectively.

Expanding (\ref{eq:B_FFT}) yields the coupled equations
\begin{align}
\mathbf{F}\mathbf{v} &= \mu \mathbf{u}, \\
\mathbf{F}^T \mathbf{u} &= \mu \mathbf{v}.
\end{align}

Substituting one equation into the other gives
\begin{align}
\mathbf{F}^T \mathbf{F} \mathbf{v} &= \mu^2 \mathbf{v}, \\
\mathbf{F} \mathbf{F}^T \mathbf{u} &= \mu^2 \mathbf{u},
\end{align}
which shows that $\mu^2$ is an eigenvalue of both $\mathbf{F}^T\mathbf{F}$ and $\mathbf{F}\mathbf{F}^T$. Consequently, for every eigenvalue $\lambda_i$ of $\mathbf{F}^T\mathbf{F}$, the adjacency matrix $\mathbf{A}$ admits a pair of eigenvalues
\begin{equation}
\mu_i = \pm \sqrt{\lambda_i}.
\end{equation}

Then, the two largest eigenvalues of $\mathbf{A}$ are given by $\mu_1 = \sqrt{\lambda_1}$ and $\mu_2 = \sqrt{\lambda_2}$. The spectral gap $\Delta$ is defined as
\begin{equation}\label{eq:App_Delta}
\Delta = \frac{\mu_1 - \mu_2}{\mu_1} = 1 - \sqrt{\frac{\lambda_2}{\lambda_1}}.
\end{equation}

Thus, we obtain
\begin{equation}\label{eq:ead_final}
N_{\text{EAD}} \geq \frac{KL}{1 + (1 - \Delta)^2 \cdot \frac{JL - V}{V}}. 
\end{equation}

Furthermore, we define the minimum EAD as
\begin{equation}\label{eq:gamma_vs_delta}
N_{\text{EAD}}^{\min}
= \frac{KL}{1 + (1 - \Delta)^2 \cdot \frac{JL - V}{V}},
\end{equation}
which shows that a larger spectral gap leads to a larger MED and therefore a lower pairwise error probability. Consequently, spatial coupling improves the error performance of SC-SCMA through the following mechanism
\begin{equation}
\Delta\uparrow
\Longrightarrow
N_{\mathrm{EAD}}\uparrow
\Longrightarrow
d_{\min}(\mathcal V)\uparrow
\Longrightarrow
\mathrm{PEP}\downarrow.
\end{equation}

\section{Low-Complexity Design of SC-SCMA}
In this section, we develop a low-complexity SCFM and codebook design framework for SC-SCMA systems. We first replace the intractable EAD optimization with a tractable spectral gap metric based on the windowed SCFM, enabling scalable design with reduced complexity. A coupling assignment matrix is introduced to systematically construct the SCFM while preserving sparsity and structural constraints.

Furthermore, we show that the system performance is dominated by the LUG, which significantly reduces the dimensionality of the codebook optimization. Based on this observation, the codebook design is reformulated over the LUG with structured constellations, achieving substantial complexity reduction while maintaining near-optimal performance.

\subsection{Low-complexity Design of SCFM}
As established in the previous sections, the EAD plays a fundamental role in determining the error performance of SC-SCMA systems and is closely related to the spectral gap of the underlying factor graph. However, directly maximizing the EAD is computationally prohibitive due to the exponential complexity $\mathcal{O}(2^{J_{\text{tot}}})$ associated with enumerating all user subsets. To address this issue, we adopt the spectral gap as a tractable metric and develop a low-complexity SCFM design framework.

It is important to emphasize that the optimization is performed on the windowed SCFM rather than the SCFM. For a coupled system with chain length $L$, computing the spectral gap of the global SCFM requires eigenvalue decomposition of a matrix with dimension proportional to $KL$, resulting in a complexity of $\mathcal{O}((KL)^3)$. In contrast, the windowed SCFM has a fixed size determined by the window length $w$, leading to a significantly reduced complexity of $\mathcal{O}((Kw)^3)$, which is independent of $L$. This makes the proposed design scalable for large systems.

The key idea of SCFM construction is to decompose the prototype matrix $\mathbf{F}_p$ into a set of sub-matrices $\{\mathbf{W}_i\}_{i=0}^{c}$ corresponding to different coupling levels. To formalize this process, we introduce a coupling assignment matrix $\mathbf{C}_{K \times J}$, where each entry $C_{k,j} \in \{0,1,\dots,c+1\}$ specifies the coupling level (i.e., block index shift) assigned to the edge between RN $k$ and user $j$. 

Based on $\mathbf{C}$, the sub-matrices can be constructed as
\begin{equation}
\mathbf{W}_i[k,j]=
\begin{cases}
1, & \text{if } \mathbf{C}[k,j]=i+1, \\
0, & \text{otherwise}.
\end{cases}
\end{equation}

In this way, the matrix $\mathbf{C}$ provides a compact representation of the spatial coupling structure, where each nonzero entry in $\mathbf{F}_p$ is mapped to a specific coupling level.

As an illustrative example, the matrices $\mathbf{W}_0$ in (\ref{eq:BF1}) and $\mathbf{W}_1$ in (\ref{eq:BF2}) can be obtained using the following coupling assignment matrix
\begin{equation}\label{eq:example_dm}
\mathbf{C}_{4\times 6} =
\begin{bmatrix}
0 & 2 & 1 & 0 & 1 & 0 \\
2 & 0 & 2 & 0 & 0 & 1 \\
0 & 1 & 0 & 1 & 0 & 2 \\
1 & 0 & 0 & 2 & 2 & 0
\end{bmatrix}.
\end{equation}

Since directly maximizing EAD is combinatorial, we instead maximize its computable lower bound through the spectral gap. Therefore, we formulate the SCFM design as the following optimization problem
\begin{subequations} \label{eq:SCM_Optimization}
\begin{align}
\max_{\mathbf{A}} \quad & \Delta(\mathbf{F}_w) \\
\text{s.t.} \quad 
& 0 \leq \mathbf{C}[k,j] \leq c+1, \\
& G_d(\mathbf{a}_j) > 1, \\
& k=1,\dots,K,\ j=1,\dots,J,
\end{align}
\end{subequations}
where $\Delta(\mathbf{F}_w)$ denotes the spectral gap of the adjacency matrix constructed from the windowed SCFM $\mathbf{F}_w$. $G_d(\mathbf{a}_j)$ counts the number of distinct nonzero elements in the $j$-th of $\mathbf{A}$. The constraint $G_d(\mathbf{a}_j) > 1$ ensures that the edges associated with each user are distributed across multiple coupling levels, which prevents all connections from collapsing into a single block. This promotes diversity and avoids severe performance degradation due to concentrated interference. If such diversity is not required, this constraint can be relaxed.

It is worth noting that the proposed formulation significantly reduces computational complexity. Instead of directly optimizing the EAD, which requires exhaustive enumeration, the spectral gap can be efficiently computed via eigenvalue decomposition. Furthermore, since the optimization is performed on the windowed SCFM, the complexity remains manageable even for large-scale systems.

\begin{figure}
\centering
\includegraphics[width=3.5in]{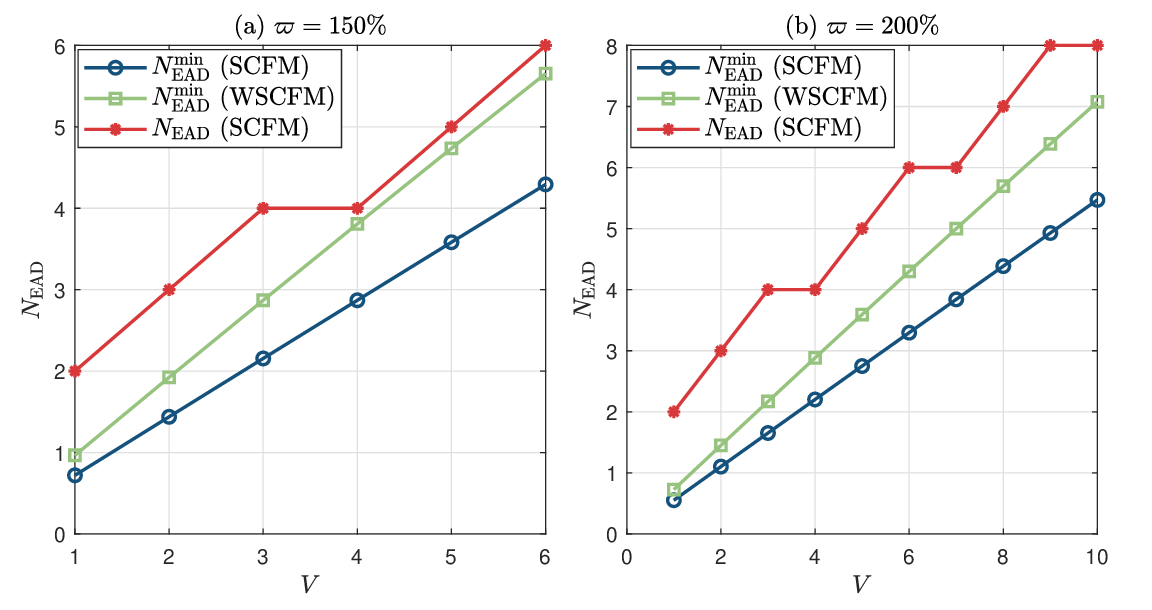} \\
\caption{Comparison of ${N}_\text{EAD}$ and ${N}_\text{EAD}^{\min}$ versus $V$ for SCFM and WSCFM with 1-level spatial coupling.}
\label{fig:gamma_min_vs_v}
\end{figure}

To validate the effectiveness of the proposed design, Fig.~\ref{fig:gamma_min_vs_v} compares the theoretical lower bound ${N}_\text{EAD}^{\min}$ and the actual ${N}_\text{EAD}$ under different overloading $\varpi$. It is observed that the windowed SCFM provides a much tighter approximation to the true EAD than the conventional SCFM. This is because the windowed structure more accurately captures the localized connectivity induced by spatial coupling, leading to a more precise characterization of the spectral properties.

\subsection{Low-Complexity Design of Prototype SCMA Codebook}
\begin{figure}
\centering
\includegraphics[width=3.5in]{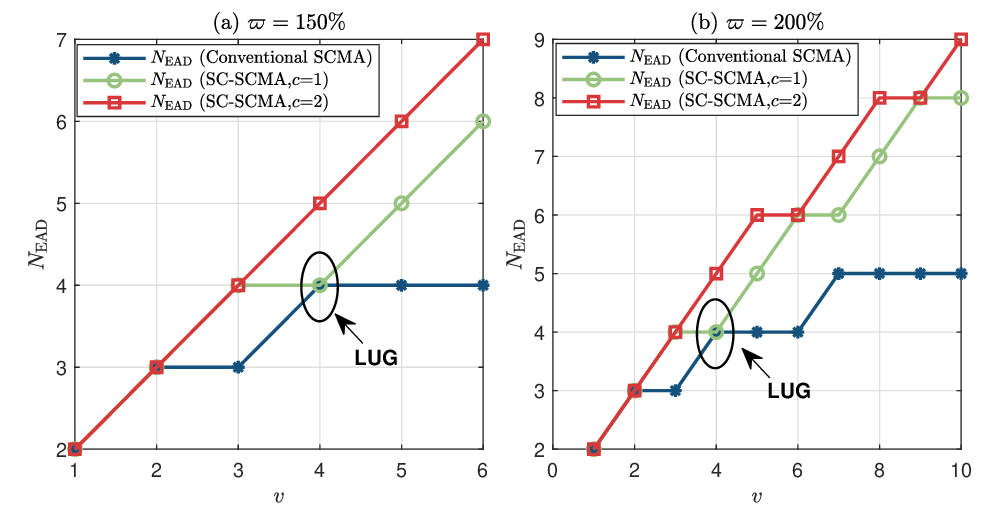} \\
\caption{Comparison of ${N}_\text{EAD}$ versus $V$ for conventional SCMA and SC-SCMA.}
\label{fig:gamma_vs_v}
\end{figure}

Fig.~\ref{fig:gamma_vs_v} compares the EAD performance of conventional SCMA and SC-SCMA. It can be observed that SC-SCMA significantly outperforms conventional SCMA due to the enlarged spectral gap induced by spatial coupling. 

Interestingly, the EAD curves of SC-SCMA and conventional SCMA coincide for certain numbers of active users. This implies that, for these specific user group sizes, spatial coupling does not provide additional performance gains. Let $V_\tau$ denote the maximum number of users for which this equality holds. We refer to such user groups as the LUG.

This observation reveals a fundamental property of SC-SCMA, where the overall system performance is primarily governed by the LUG, which constitutes the dominant bottleneck in multi-user interference suppression. When the user-group cardinality exceeds $V_{\tau}$, spatial coupling introduces enhanced inter-block connectivity, enabling global information exchange over the extended factor graph. Consequently, the multi-user interference is progressively averaged over a larger number of resource dimensions, and the system performance approaches the single-user bound.

Based on this insight, the codebook design problem can be significantly simplified. Instead of optimizing the MED over all $J$ users, it is sufficient to focus only on the LUG. This reduces the optimization complexity from $\mathcal{O}(M^{2J})$ to $\mathcal{O}(M^{2V_\tau})$, leading to a dramatic reduction in computational burden. Considering $M=4$ and $J=6$, the complexity is reduced from $4^{12}$ and $4^{20}$ to $4^8$ for $\varpi=150\%$ and $\varpi=200\%$, corresponding to reductions of 99.61\% and 99.99\%, respectively. This demonstrates that the LUG-based design provides an efficient way to handle high-dimensional codebook optimization while preserving near-optimal performance.

While conventional SCMA codebook design approaches remain applicable, the key distinction in SC-SCMA lies in shifting the optimization objective from global user sets to the LUG. To balance design flexibility and performance, we adopt the structured codebook framework proposed in \cite{liDesignPowerImbalancedSCMA2022}.

Specifically, we first define a one-dimensional constellation set of size $M$ as
\begin{equation}
    \mathbf{\mathcal{A}}^{M}=\{\mathbf{\mathcal{A}}^{M}_1,\mathbf{\mathcal{A}}^{M}_2,\cdots,\mathbf{\mathcal{A}}^{M}_{d_v}\},
\end{equation}
where $d_v$ denotes the number of nonzero elements in each SCMA codeword.

Each vector $\mathbf{\mathcal{A}}^{M}_i$ follows a symmetric structure
\begin{equation}
    \mathbf{\mathcal{A}}^{M}_i = [a_{\frac{M}{2}},\cdots,a_1,-a_1,\cdots,-a_{\frac{M}{2}}],
\end{equation}
where the amplitudes are defined as $a_k=(k-1)(\rho_i-1)+1$, with $\rho_i$ being a design parameter.

The multidimensional constellation is constructed by replicating $\mathbf{\mathcal{A}}^{M}$ across $d_v$ dimensions and applying permutation mappings
\begin{equation}
    \mathbf{\pi}_n:[1,\cdots,M]\rightarrow[\pi_{n,1},\cdots,\pi_{n,M}],
\end{equation}
to enhance diversity and improve distance properties.

In this work, we introduce a signature matrix $\mathbf{\mathcal{I}}_{K\times J}$ to construct all sparse codebooks for SCMA systems. The proposed matrix, exemplified by the $(4\times 6)$ and $(5\times 10)$ case
\begin{equation}
   \mathbf{\mathcal{I}}_{4\times 6}=\begin{bmatrix}
        0&\mathcal{I}_1^1&\mathcal{I}_2^1&0&\mathcal{I}_3^1&0 \\
        \mathcal{I}_1^1&0&\mathcal{I}_2^2&0&0&\mathcal{I}_3^1 \\
        0&\mathcal{I}_3^2&0&\mathcal{I}_2^1&0&\mathcal{I}_1^2 \\
        \mathcal{I}_3^2&0&0&\mathcal{I}_2^2&\mathcal{I}_1^2&0
    \end{bmatrix}, 
\end{equation}
\begin{equation}
\begin{split}
&\mathbf{\mathcal{I}}_{5\times 10}\\&=\begin{bmatrix}
        \mathcal{I}_1^1&\mathcal{I}_2^1&\mathcal{I}_3^1&\mathcal{I}_4^1&0&0&0&0&0&0 \\
        \mathcal{I}_4^2&0&0&0&\mathcal{I}_1^1&\mathcal{I}_2^1&\mathcal{I}_3^1&0&0&0 \\
        0&\mathcal{I}_3^1&0&0&\mathcal{I}_4^2&0&0&\mathcal{I}_1^1&\mathcal{I}_2^1&0 \\
        0&0&\mathcal{I}_2^2&0&0&\mathcal{I}_3^2&0&\mathcal{I}_4^2&0&\mathcal{I}_1^1 \\
        0&0&0&\mathcal{I}_1^2&0&0&\mathcal{I}_2^2&0&\mathcal{I}_3^2&\mathcal{I}_4^2 \\
    \end{bmatrix}, \\
\end{split}
\end{equation}
where $\mathbf{\mathcal{I}}$ maps each non-zero entry $\mathcal{I}_p^q,1\leq p\leq d_f,1\leq q\leq d_v$ to a row vector $\xi_p\exp(j\theta_p)\mathbf{\pi}_q(\mathbf{\mathcal{A}}_p)^T$ with scaling operation $\xi_p$ and phase rotation operation $\exp(j\theta_p)$. Zero entries in $\mathbf{\mathcal{I}}$ correspond to null vectors $\mathbf{0}_M^T$, enforcing sparsity. 
This structure enables efficient derivation of user-specific codebooks, e.g., 
\begin{equation}
\begin{split}
   \mathbf{\chi}_1=[\mathbf{0}_{M}^T,&\xi_1\exp(j\theta_1)\mathbf{\pi}_1(\mathbf{\mathcal{A}}_1)^T, \\
   &\mathbf{0}_{M}^T,\xi_3\exp(j\theta_2)\mathbf{\pi}_2(\mathbf{\mathcal{A}}_3)^T]^T.  \\
\end{split}
\end{equation}

Finally, the codebook design over Gaussian channels is formulated as
\begin{subequations} \label{eq:Codebook Optimization}
\begin{align}
\max_{\mathbf{\xi},\mathbf{\rho},\mathbf{\theta}} \quad 
& d_{\min}\left( \mathbf{\chi}^{\mathrm{LUG}} \right) \\
\text{s.t.} \quad
& \| \mathbf{\chi} \|_2^2 = MV, \\
& 1< \xi_i \leq 6,\ i=1,\cdots,d_f, \\
& 3\leq \rho_i \leq 6,\ i=1,\cdots,d_f, \\
& 0\leq \theta_i \leq \pi,\ i=1,\cdots,d_f,
\end{align}
\end{subequations}
where $\mathbf{\chi}^{\mathrm{LUG}}$ represents the corresponding codebooks for the LUG.

In this work, the optimization problem is solved using genetic algorithms, which provide an effective means of exploring the high-dimensional parameter space and identifying near-optimal solutions.

\section{Numerical Results}
In this section, we provide comprehensive simulation results to evaluate the performance of the proposed OFDM-based SC-SCMA system across various propagation environments. The detailed simulation parameters are summarized in Table \ref{table_sim_params}. The system operates at a sampling rate of $15.36$ MHz with an FFT size of $512$ and a cyclic prefix (CP) length of $64$, resulting in a subcarrier spacing of $30$ kHz. To ensure statistical significance, each BER curve is obtained by averaging over $10^3$ independent frames. To evaluate receiver robustness, we consider three distinct channel scenarios based on the 3GPP TR 38.901 specification. First, the AWGN channel is employed as a theoretical benchmark to establish the performance upper bound. Second, the TDL-C (NLOS) channel is utilized to simulate a typical urban macro-cell environment characterized by a rich scattering profile. It consists of $24$ taps with a nominal root-mean-square (RMS) delay spread of $\tau_{\mathrm{rms}} = 300$ ns, inducing significant frequency-selective fading. Finally, the TDL-D (LOS) channel represents a line-of-sight scenario with a smaller delay spread of $\tau_{\mathrm{rms}} = 30$ ns. In this model, the first tap incorporates a dominant LOS component with a Rician $K$-factor of $13.3$ dB, while the remaining $13$ taps follow Rayleigh fading distributions. The receiver performs joint detection using the MPA scheme, with the maximum number of iterations fixed at $T_{\max} = 6$.
\begin{table}[htbp]
\renewcommand{\arraystretch}{1.2}
\setlength{\tabcolsep}{4pt}
\caption{Simulation parameters for OFDM-SCMA system}
\label{table_sim_params}
\centering
\begin{tabular}{lccc}
\toprule
\textbf{Parameters} & \textbf{AWGN} & \textbf{TDL-C} & \textbf{TDL-D} \\
\midrule
Channel Scenario & Noise Only & Non-Line-of-Sight & Line-of-Sight \\
RMS Delay Spread & -- & $300$ ns & $30$ ns \\
Number of Taps & -- & $24$ & $14$ \\
Rician $K$-factor & -- & $0$ & $13.3$ dB \\
Sampling Rate & $15.36$ MHz & $15.36$ MHz & $15.36$ MHz \\
FFT Size& $512$& $512$& $512$\\
CP Length & $64$ & $64$ & $64$ \\
Subcarrier Spacing  & $30$ kHz & $30$ kHz & $30$ kHz \\
Decoder Type & MPA & MPA & MPA \\
Number of Iterations & 6 & 6 & 6 \\
\bottomrule
\end{tabular}
\end{table}

The Huang codebook \cite{huangDownlinkSCMACodebook2022} and the Luo codebook \cite{luoDesignLowProjectionSCMA2023} are adopted since they achieve the best BER performance with $\varpi=150\%$ and $\varpi=200\%$, respectively, over the AWGN channel and therefore serve as appropriate benchmark schemes. By contrast, the Huawei codebook \cite{altera_scma_env1} is selected due to its relatively inferior BER performance under the same channel condition, which enables a more explicit evaluation of the performance gains brought by SC-SCMA.

\begin{figure}
\centering
\includegraphics[width=3.5in]{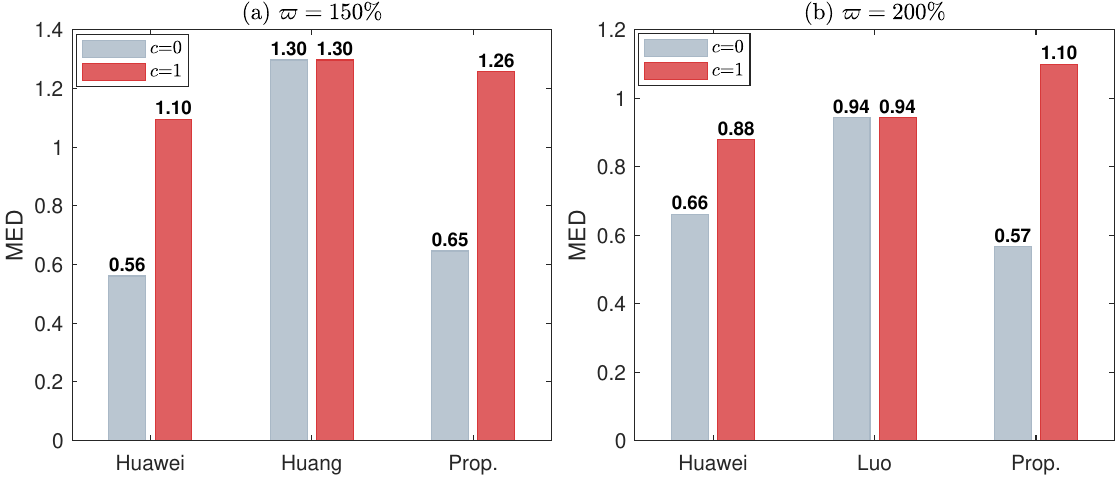} \\
\caption{The MED comparison for different codebooks before and after spatial coupling with $\varpi=150\%$ and $\varpi=200\%$.}
\label{fig:improved_med_mbpd}
\end{figure}

Fig. \ref{fig:improved_med_mbpd} illustrates the MED variations of different codebooks before and after the application of spatial coupling, corresponding to c=0 and c=1, respectively. As shown in Fig. \ref{fig:improved_med_mbpd}(a), both the Huawei codebook and the proposed codebook experience substantial MED improvement. In particular, the MED of the proposed codebook increases markedly from 0.65 to 1.26, approaching the currently known maximum MED value of 1.30, which corresponds to an improvement of approximately 94\%. In addition, the optimization complexity of the proposed codebook is reduced by 99.61\% compared with existing codebooks. The MED of the Huang codebook remains unchanged after spatial coupling since its MED lies within the user threshold given by $V_{\tau}=1<4$ as discussed earlier. Fig. \ref{fig:improved_med_mbpd}(b) reports the results for an $\mathbf{F}_{5\times 10}$ system with an overloading factor $\varpi$ of 200\%. In this case, the MED is evaluated for only 6 of the 10 users because direct computation for all users is infeasible. The results indicate that both the Huawei codebook and the proposed codebook still achieve considerable MED gains. Specifically, the proposed codebook exhibits an increase in the minimum MED from 0.57 to 1.10, corresponding to a gain of nearly 93\%, while the optimization complexity is reduced by 99.99\%. The MED of the Luo codebook remains invariant under spatial coupling, which can be attributed to the LUG effect.

\begin{figure*}
\centering
\includegraphics[width=6.5in]{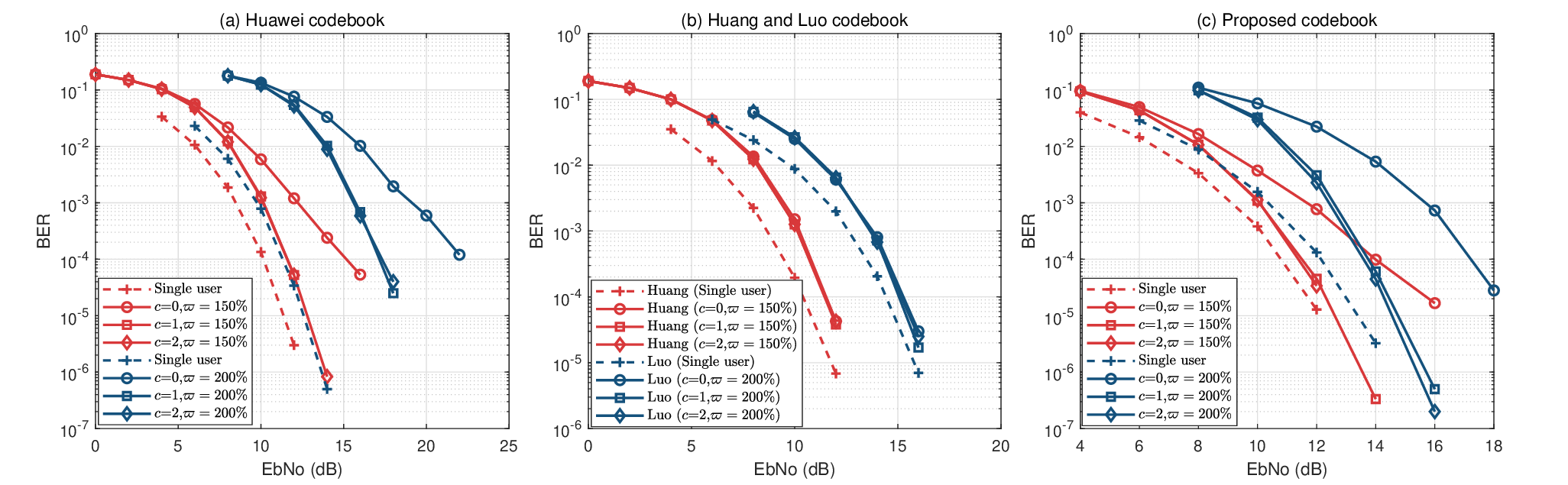} \\
\caption{BER performance comparison for different codebooks with $\varpi=150\%$ and $\varpi=200\%$ over AWGN channel.}
\label{fig:ber_for_f4x6}
\end{figure*}

Fig. \ref{fig:ber_for_f4x6} presents the BER performance of various codebooks under the AWGN channel with $\varpi \in \{150\%, 200\%\}$ and $c \in \{0, 1, 2\}$. The simulation results demonstrate that while the proposed codebook may not exhibit the best initial performance in uncoupled scenarios, it possesses stronger
structural compatibility with spatial coupling, enabling it to approach the single-user bound more effectively than existing benchmarks. As illustrated in Fig. \ref{fig:ber_for_f4x6}, the BER performance of the proposed codebook improves significantly as the coupling level increases from $c=0$ to $c=1$, with the performance saturating at $c=2$, indicating that the maximum spatial coupling gain is attained early. The performance improvement is consistent across different overloading factors. At $\varpi=150\%$, the uncoupled proposed codebook ($c=0$) shows a 3 dB gap from the single-user bound at a BER of $10^{-4}$; however, this gap is drastically narrowed to merely 0.5 dB after spatial coupling ($c \geq 1$). Even under the more congested scenario of $\varpi=200\%$, where the proposed codebook initially lags behind the single-user bound by 5 dB, the spatial coupling mechanism recovers the performance to within 1.7 dB of the bound. In contrast, the benchmark codebooks exhibit limited gains from spatial coupling. As shown in Fig. \ref{fig:ber_for_f4x6}, the Huang and Luo codebooks initially outperform the proposed codebook by 2.5 dB at $c=0$, yet its BER curves remain nearly stationary regardless of the coupling level. This lack of improvement stems from their MEDs being associated with only a single user, which fails to exceed the LUG  threshold required to trigger spatial coupling gains. Consequently, the proposed codebook achieves a significant competitive advantage in coupled systems, maintaining a steeper waterfall curve and superior power efficiency under heavy overloading.

\begin{figure}
\centering
\includegraphics[width=3.5in]{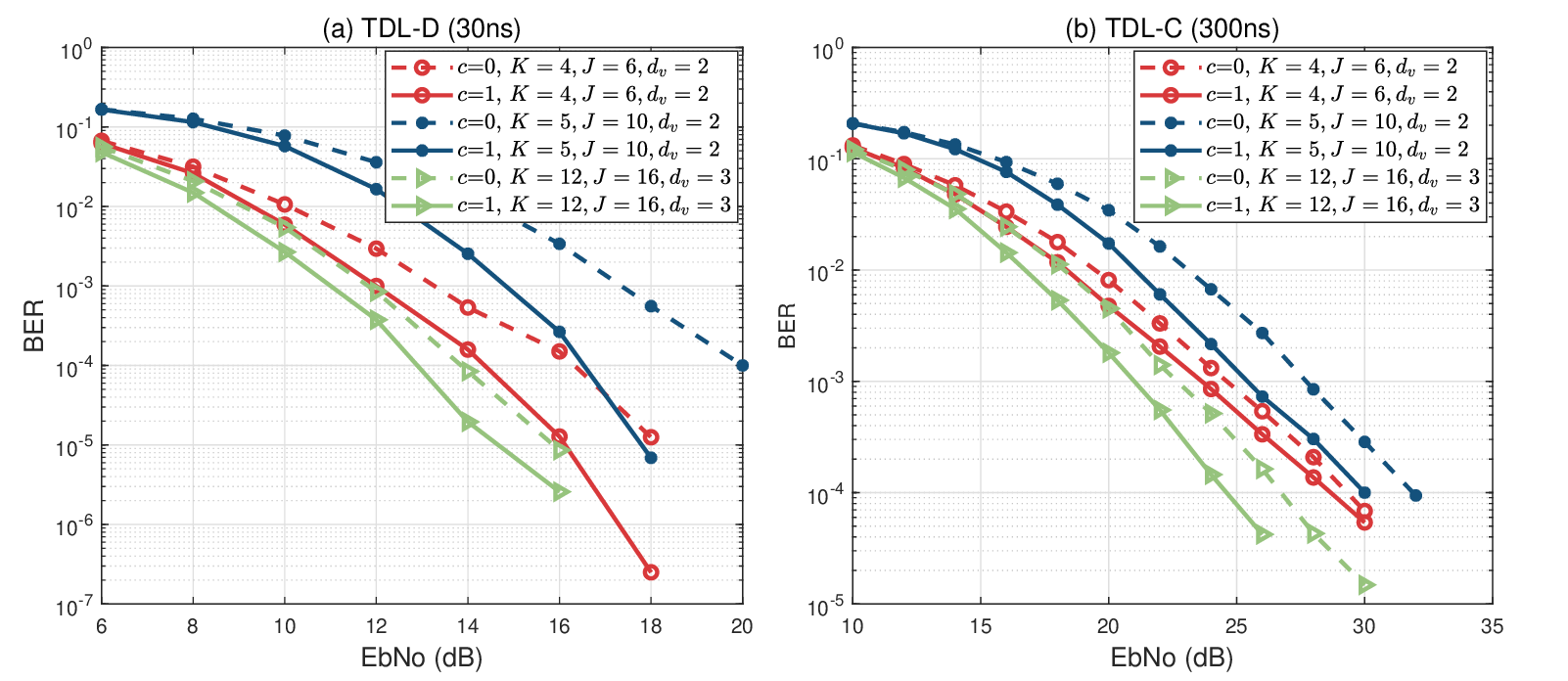} \\
\caption{BER performance of the proposed codebook in OFDM systems over TDL-C and TDL-D channels.}
\label{fig:ber_for_f4x6_ofdm}
\end{figure}

Fig.~\ref{fig:ber_for_f4x6_ofdm} presents the BER performance of the proposed codebooks in an OFDM-based system over realistic frequency-selective fading channels, namely the TDL-D channel with a 30 ns delay spread and the TDL-C channel with a 300 ns delay spread. Three prototype matrices, $(4\times6,d_v=2)$, $(5\times10,d_v=2)$, and $(12\times16,d_v=3)$, are considered with coupling levels $c\in\{0,1\}$. Overall, the results demonstrate that spatial coupling remains highly effective in multipath fading environments, consistently enhancing the BER performance across different overloading factors and codebook structures.

Specifically, for the TDL-D channel shown in Fig.~\ref{fig:ber_for_f4x6_ofdm}(a), the spatially coupled designs ($c=1$) achieve SNR gains of approximately 2 dB, 3.7 dB, and 1.2 dB at a BER of $10^{-4}$ for the $(4\times6,d_v=2)$, $(5\times10,d_v=2)$, and $(12\times16,d_v=3)$ codebooks, respectively. Among the considered schemes, the $(5\times10,d_v=2)$ codebook exhibits the largest gain, indicating that the effectiveness of spatial coupling becomes more pronounced under higher overloading factors. Although the $(12\times16,d_v=3)$ codebook achieves a comparatively smaller coupling gain, it consistently attains the lowest BER over the investigated SNR range, demonstrating its superior overall error-rate performance.

For the more challenging TDL-C channel shown in Fig.~\ref{fig:ber_for_f4x6_ofdm}(b), the required SNR increases due to the stronger frequency selectivity introduced by the larger delay spread. Nevertheless, spatial coupling continues to provide notable performance improvements, yielding gains of approximately 0.8 dB, 2 dB, and 2 dB at a BER of $10^{-4}$ for the three codebooks, respectively. These results confirm that the proposed spatially coupled codebooks maintain strong robustness against severe multipath fading.

\begin{figure*}
\centering
\includegraphics[width=6.5in]{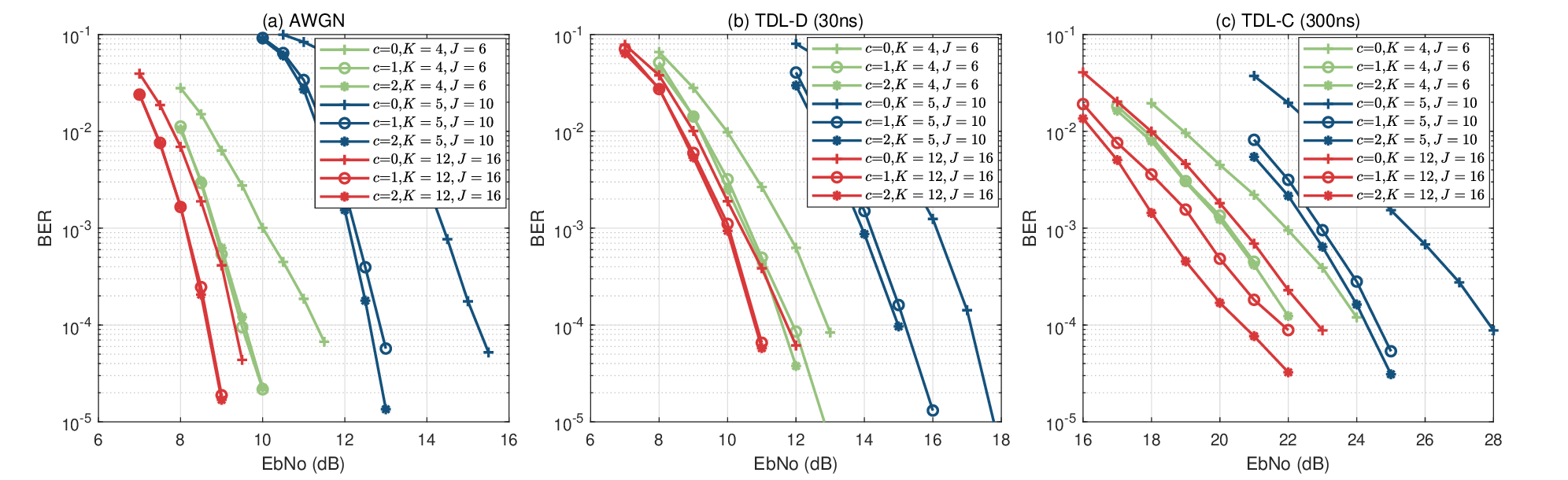} \\
\caption{LDPC coded BER performance of proposed codebook ($R=5/6$).}
\label{fig:ber_ldpc}
\end{figure*}

Fig.~\ref{fig:ber_ldpc} presents the coded BER performance of the proposed codebook with an LDPC code of rate $5/6$ under AWGN, TDL-D (30 ns), and TDL-C (300 ns) channels. It can be observed that spatial coupling consistently improves the BER performance for all considered codebook configurations and channel conditions, with the proposed $c=2$ design achieving the best performance. For the AWGN channel in Fig.~\ref{fig:ber_ldpc}(a), the gain introduced by spatial coupling increases with the codebook dimension. At BER $=10^{-4}$, the largest gain of 2.6 dB is achieved by the $(K,J)=(5,10)$ codebook, while the smallest gain of 0.6 dB is observed for $(K,J)=(12,16)$. The $(K,J)=(4,6)$ configuration provides an intermediate gain of 1.8 dB. This trend suggests that the codebooks with larger overloading benefit more from the proposed spatial coupling mechanism.

A similar behavior can be observed for the TDL-D channel in Fig.~\ref{fig:ber_ldpc}(b). Despite the performance degradation caused by frequency-selective fading, the proposed spatially coupled codebook maintains a clear advantage over its uncoupled counterpart. At BER $=10^{-4}$, the gains achieved by the $(K,J)=(4,6)$, $(5,10)$, and $(12,16)$ codebooks are 1.5 dB, 2.1 dB, and 0.8 dB, respectively. Among them, the $(5,10)$ codebook achieves the largest gain, whereas the $(12,16)$ codebook exhibits the smallest improvement.

The superiority of the proposed design remains evident in the more challenging TDL-C channel shown in Fig.~\ref{fig:ber_ldpc}(c), where the larger delay spread results in more severe channel dispersion. At BER $=10^{-4}$, the gains provided by spatial coupling are 1.8 dB, 3.8 dB, and 2.2 dB for the $(4,6)$, $(5,10)$, and $(12,16)$ codebooks, respectively. Similar to the previous cases, the largest gain is achieved by the $(5,10)$ codebook, while the smallest gain is obtained for the $(4,6)$ configuration. Notably, for the $(12,16)$ codebook with $d_v=3$, increasing the coupling depth from $c=1$ to $c=2$ still yields an additional gain of approximately $1.2$ dB at BER $=10^{-4}$. This result indicates that stronger spatial coupling can further improve the convergence behavior of the iterative receiver when the codebook possesses a higher node degree and richer connectivity structure.

\section{Conclusion}

This paper proposed an SC-SCMA framework to address the scalability limitations of conventional SCMA. By introducing structured spatial coupling across multiple SCMA blocks, the proposed scheme transforms the block-diagonal factor graph into a globally connected structure, enabling enhanced user--resource interaction while preserving sparsity and low-complexity MPA detection.

A unified PEP-based analytical framework was developed to characterize the performance gain of SC-SCMA. The results showed that spatial coupling enlarges the EAD, which leads to a strict improvement in the MED of the superimposed constellation. By leveraging spectral graph theory, a direct relationship between the spectral gap and a lower bound on the EAD was established, providing a computable structural metric that links graph connectivity to distance performance.

Building upon these insights, a low-complexity codebook design approach was developed. The SCFM design was reformulated as a spectral-gap maximization problem, and the windowed SCFM was introduced to enable efficient evaluation in large-scale systems. Furthermore, by exploiting the LUG property, the codebook design was reduced from a global high-dimensional optimization to a localized optimization over dominant user groups, significantly reducing complexity while preserving MED performance.

Simulation results over AWGN and frequency-selective fading channels verified the analysis and demonstrated that SC-SCMA achieves consistent MED and BER gains under high overloading conditions. These results confirm that SC-SCMA provides a scalable and structure-aware design paradigm for SCMA systems. The PEP framework was developed for the AWGN channel, where the minimum Euclidean distance governs performance. An extension to Rayleigh fading channels shows that the EAD controls both the diversity order and the minimum product distance. Codebook optimisation targeting the minimum product distance for fading channels is left as a future work.

\appendices
\section{Proof of the Dominant Eigenvalue} \label{app:a}

The dominant eigenvalue $\lambda_1$ of the matrix $F^T F$ is given by 
$\lambda_1 = d_f d_v$, with $e_1 = \frac{1}{\sqrt{J_{\mathrm{tot}}}} \mathbf{1}_{J_{\mathrm{tot}}}$ being 
the associated eigenvector. This can be justified as follows

\begin{equation}
    \mathbf{F}^T\mathbf{F}\mathbf{e}_1=\mathbf{F}^T(\mathbf{F}\mathbf{e}_1)=\frac{d_f}{\sqrt{J_{\mathrm{tot}}}}\mathbf{F}^T\mathbf{1}_{K_{\mathrm{tot}}}=d_fd_v\mathbf{e}_1,
\end{equation}
where $d_fd_v$ is the eigenvalue of $\mathbf{F}^T\mathbf{F}$ and which implies $\lambda_1\geq d_fd_v$.

\begin{equation}
\begin{split}
    \lambda_1&=\max_{\left \| \mathbf{x} \right \|_2^2=1}\left \| \mathbf{F}\mathbf{x} \right \|_2^2 \\
    &=\max_{\left \| \mathbf{x} \right \|_2^2=1}\sum_{i=1}^{K_{\mathrm{tot}}}(\sum_{j=1}^{J_{\mathrm{tot}}}\mathbf{F}_{i,j}x_j)^2 \\
    &\leq \max_{\left \| \mathbf{x} \right \|_2^2=1}\sum_{i=1}^{K_{\mathrm{tot}}}(\sum_{j=1}^{J_{\mathrm{tot}}}\mathbf{F}_{i,j})^2\left \| \mathbf{x} \right \|_2^2 \\
    &=\left \| \mathbf{F} \right \|_2^2=d_fd_v,
\end{split} 
\end{equation}
where $\lambda_1\leq d_fd_v$ and we have already proven $\lambda_1\geq d_fd_v$, it necessarily follows that $\lambda_1=d_fd_v$.

\bibliographystyle{IEEEtran}
\bibliography{references}

@article{liuNonorthogonal2017,
  title = {Nonorthogonal Multiple Access for 5G and Beyond},
  author = {Liu, Yuanwei and Qin, Zhijin and Elkashlan, Maged and Ding, Zhiguo and Nallanathan, Arumugam and Hanzo, Lajos},
  year = 2017,
  month = dec,
  journal = {Proc. IEEE},
  volume = {105},
  number = {12},
  pages = {2347--2381},
  issn = {0018-9219, 1558-2256},
  doi = {10.1109/JPROC.2017.2768666},
  urldate = {2026-02-20},
  copyright = {https://creativecommons.org/licenses/by/3.0/legalcode},
  langid = {english}
}

@article{chaturvediTutorialDecodingTechniques2022,
  title = {A {{Tutorial}} on {{Decoding Techniques}} of {{Sparse Code Multiple Access}}},
  author = {Chaturvedi, Saumya and Liu, Zilong and Bohara, Vivek Ashok and Srivastava, Anand and Xiao, Pei},
  year = 2022,
  journal = {IEEE Access},
  volume = {10},
  pages = {58503--58524},
  issn = {2169-3536},
  doi = {10.1109/ACCESS.2022.3178127},
  urldate = {2026-02-20},
  copyright = {https://creativecommons.org/licenses/by-nc-nd/4.0/},
  langid = {english}
}

@article{hoshyarNovelLowDensitySignature2008,
  title = {Novel {{Low-Density Signature}} for {{Synchronous CDMA Systems Over AWGN Channel}}},
  author = {Hoshyar, Reza and Wathan, Ferry P. and Tafazolli, Rahim},
  year = 2008,
  month = apr,
  journal = {IEEE Trans. Signal Process.},
  volume = {56},
  number = {4},
  pages = {1616--1626},
  issn = {1053-587X},
  doi = {10.1109/TSP.2007.909320},
  urldate = {2026-02-20},
  copyright = {https://ieeexplore.ieee.org/Xplorehelp/downloads/license-information/IEEE.html},
  langid = {english}
}

@article{huangDownlinkSCMACodebook2022,
  title = {Downlink {{SCMA Codebook Design With Low Error Rate}} by {{Maximizing Minimum Euclidean Distance}} of {{Superimposed Codewords}}},
  author = {Huang, Chinwei and Su, Borching and Lin, Tingyi and Huang, Yenming},
  year = 2022,
  month = may,
  journal = {IEEE Trans. Veh. Technol.},
  volume = {71},
  number = {5},
  pages = {5231--5245},
  issn = {0018-9545, 1939-9359},
  doi = {10.1109/TVT.2022.3155627},
  urldate = {2026-02-20},
  copyright = {https://ieeexplore.ieee.org/Xplorehelp/downloads/license-information/IEEE.html},
  langid = {english}
}

@inproceedings{kudekarEffectSpatialCoupling2010,
  title = {The Effect of Spatial Coupling on Compressive Sensing},
  booktitle = {2010 48th {{Annual Allerton Conference}} on {{Communication}}, {{Control}}, and {{Computing}} ({{Allerton}})},
  author = {Kudekar, Shrinivas and Pfister, Henry D.},
  year = 2010,
  month = sep,
  pages = {347--353},
  publisher = {IEEE},
  address = {Monticello, IL, USA},
  doi = {10.1109/ALLERTON.2010.5706927},
  urldate = {2026-02-20},
  isbn = {978-1-4244-8215-3},
  langid = {english}
}

@inproceedings{kudekarThresholdSaturationBMS2010,
  title = {Threshold Saturation on {{BMS}} Channels via Spatial Coupling},
  booktitle = {2010 6th {{International Symposium}} on {{Turbo Codes}} \& {{Iterative Information Processing}}},
  author = {Kudekar, Shrinivas and Meassony, Cyril and Richardson, Tom and Urbankez, Rudiger},
  year = 2010,
  month = sep,
  pages = {309--313},
  publisher = {IEEE},
  address = {Brest},
  doi = {10.1109/ISTC.2010.5613887},
  urldate = {2026-02-20},
  isbn = {978-1-4244-6744-0 978-1-4244-6746-4},
  langid = {english}
}

@article{kudekarThresholdSaturationSpatial2011,
  title = {Threshold {{Saturation}} via {{Spatial Coupling}}: {{Why Convolutional LDPC Ensembles Perform So Well}} over the {{BEC}}},
  shorttitle = {Threshold {{Saturation}} via {{Spatial Coupling}}},
  author = {Kudekar, Shrinivas and Richardson, Thomas J. and Urbanke, R{\"u}diger L.},
  year = 2011,
  month = feb,
  journal = {IEEE Trans. Inform. Theory},
  volume = {57},
  number = {2},
  pages = {803--834},
  issn = {0018-9448, 1557-9654},
  doi = {10.1109/TIT.2010.2095072},
  urldate = {2026-02-20},
  copyright = {https://ieeexplore.ieee.org/Xplorehelp/downloads/license-information/IEEE.html},
  langid = {english}
}

@article{liDesignPowerImbalancedSCMA2022,
  title = {Design of {{Power-Imbalanced SCMA Codebook}}},
  author = {Li, Xudong and Gao, Zhicheng and Gui, Yiming and Liu, Zilong and Xiao, Pei and Yu, Lisu},
  year = 2022,
  month = feb,
  journal = {IEEE Trans. Veh. Technol.},
  volume = {71},
  number = {2},
  pages = {2140--2145},
  issn = {0018-9545, 1939-9359},
  doi = {10.1109/TVT.2021.3132698},
  urldate = {2026-02-20},
  copyright = {https://ieeexplore.ieee.org/Xplorehelp/downloads/license-information/IEEE.html},
  langid = {english}
}

@article{liuEvolutionNOMANext2022,
  title = {Evolution of {{NOMA Toward Next Generation Multiple Access}} ({{NGMA}}) for {{6G}}},
  author = {Liu, Yuanwei and Zhang, Shuowen and Mu, Xidong and Ding, Zhiguo and Schober, Robert and {Al-Dhahir}, Naofal and Hossain, Ekram and Shen, Xuemin},
  year = 2022,
  month = apr,
  journal = {IEEE J. Select. Areas Commun.},
  volume = {40},
  number = {4},
  pages = {1037--1071},
  issn = {0733-8716, 1558-0008},
  doi = {10.1109/JSAC.2022.3145234},
  urldate = {2026-02-20},
  copyright = {https://ieeexplore.ieee.org/Xplorehelp/downloads/license-information/IEEE.html},
  langid = {english}
}

@article{liuSparseDenseComparative2021,
  title = {Sparse or {{Dense}}: {{A Comparative Study}} of {{Code-Domain NOMA Systems}}},
  shorttitle = {Sparse or {{Dense}}},
  author = {Liu, Zilong and Yang, Lie-Liang},
  year = 2021,
  month = aug,
  journal = {IEEE Trans. Wireless Commun.},
  volume = {20},
  number = {8},
  pages = {4768--4780},
  issn = {1536-1276, 1558-2248},
  doi = {10.1109/TWC.2021.3062235},
  urldate = {2026-02-20},
  copyright = {https://ieeexplore.ieee.org/Xplorehelp/downloads/license-information/IEEE.html},
  langid = {english}
}

@article{luoDesignLowProjectionSCMA2023,
  title = {A {{Design}} of {{Low-Projection SCMA Codebooks}} for {{Ultra-Low Decoding Complexity}} in {{Downlink IoT Networks}}},
  author = {Luo, Qu and Liu, Zilong and Chen, Gaojie and Xiao, Pei and Ma, Yi and Maaref, Amine},
  year = 2023,
  month = oct,
  journal = {IEEE Trans. Wireless Commun.},
  volume = {22},
  number = {10},
  pages = {6608--6623},
  issn = {1536-1276, 1558-2248},
  doi = {10.1109/TWC.2023.3244868},
  urldate = {2026-02-20},
  copyright = {https://ieeexplore.ieee.org/Xplorehelp/downloads/license-information/IEEE.html},
  langid = {english}
}

@article{hanEnablingHighOrder2018,
  title = {Enabling {{High Order SCMA Systems}} in {{Downlink Scenarios With}} a {{Serial Coding Scheme}}},
  author = {Han, Yuxi and Zhou, Wuyang and Zhao, Ming and Zhou, Shengli},
  year = 2018,
  journal = {IEEE Access},
  volume = {6},
  pages = {33796--33809},
  issn = {2169-3536},
  doi = {10.1109/ACCESS.2018.2842233},
  urldate = {2026-02-26},
  copyright = {https://ieeexplore.ieee.org/Xplorehelp/downloads/license-information/OAPA.html},
  langid = {english}
}

@inproceedings{bayestehLowComplexityTechniques2015,
  title = {Low {{Complexity Techniques}} for {{SCMA Detection}}},
  booktitle = {2015 {{IEEE Globecom Workshops}} ({{GC Wkshps}})},
  author = {Bayesteh, Alireza and Nikopour, Hosein and Taherzadeh, Mahmoud and Baligh, Hadi and Ma, Jianglei},
  year = 2015,
  month = dec,
  pages = {1--6},
  publisher = {IEEE},
  address = {San Diego, CA, USA},
  doi = {10.1109/GLOCOMW.2015.7414184},
  urldate = {2026-02-26},
  isbn = {978-1-4673-9526-7},
  langid = {english}
}

@article{dongEfficientSCMACodebook2018,
  title = {An {{Efficient SCMA Codebook Optimization Algorithm Based}} on {{Mutual Information Maximization}}},
  author = {Dong, Chao and Gao, Guili and Niu, Kai and Lin, Jiaru},
  editor = {Ansari, Imran S.},
  year = 2018,
  month = jan,
  journal = {Wireless Communications and Mobile Computing},
  volume = {2018},
  number = {1},
  pages = {8910907},
  issn = {1530-8669, 1530-8677},
  doi = {10.1155/2018/8910907},
  urldate = {2026-02-26},
  langid = {english}
}

@article{maBitInterleavedCodedSCMA,
  title = {Bit-{{Interleaved Coded SCMA With Iterative Multiuser Detection}}: {{Multidimensional Constellations Design}}},
  author = {Jinchen, Bao and Ma, Zheng and Xiao, Ming and Tsiftsis, Theodoros A and Zhongliang, Zhu},
  year = 2018,
  month = nov,
  volume = {66},
  number = {11},
  pages = {5292--5364},
  publisher = {IEEE},
  langid = {english}
}

@article{luoEnhancingSignalSpace2024,
  title = {Enhancing {{Signal Space Diversity}} for {{SCMA Over Rayleigh Fading Channels}}},
  author = {Luo, Qu and Liu, Zilong and Chen, Gaojie and Xiao, Pei},
  year = 2024,
  month = apr,
  journal = {IEEE Trans. Wireless Commun.},
  volume = {23},
  number = {4},
  pages = {3676--3690},
  issn = {1536-1276, 1558-2248},
  doi = {10.1109/TWC.2023.3310000},
  urldate = {2026-02-20},
  copyright = {https://ieeexplore.ieee.org/Xplorehelp/downloads/license-information/IEEE.html},
  langid = {english}
}

@article{mitchellSpatiallyCoupledLDPC2015,
  title = {Spatially {{Coupled LDPC Codes Constructed From Protographs}}},
  author = {Mitchell, David G. M. and Lentmaier, Michael and Costello, Daniel J.},
  year = 2015,
  month = sep,
  journal = {IEEE Trans. Inform. Theory},
  volume = {61},
  number = {9},
  pages = {4866--4889},
  issn = {0018-9448, 1557-9654},
  doi = {10.1109/TIT.2015.2453267},
  urldate = {2026-02-20},
  copyright = {https://ieeexplore.ieee.org/Xplorehelp/downloads/license-information/IEEE.html},
  langid = {english}
}

@inproceedings{nikopourSparseCodeMultiple2013,
  title = {Sparse Code Multiple Access},
  booktitle = {2013 {{IEEE}} 24th {{Annual International Symposium}} on {{Personal}}, {{Indoor}}, and {{Mobile Radio Communications}} ({{PIMRC}})},
  author = {Nikopour, Hosein and Baligh, Hadi},
  year = 2013,
  month = sep,
  pages = {332--336},
  publisher = {IEEE},
  address = {London},
  doi = {10.1109/PIMRC.2013.6666156},
  urldate = {2026-02-20},
  isbn = {978-1-4673-6235-1}
}

@article{schmalenCombiningSpatiallyCoupled,
  title = {Combining {{Spatially Coupled LDPC Codes}} with {{Modulation}} and {{Detection}}},
  author = {Schmalen, Laurent and Laboratories, Bell and Schmalen, Laurent},
  langid = {english}
}

@inproceedings{taherzadehSCMACodebookDesign2014,
  title = {{{SCMA Codebook Design}}},
  booktitle = {2014 {{IEEE}} 80th {{Vehicular Technology Conference}} ({{VTC2014-Fall}})},
  author = {Taherzadeh, Mahmoud and Nikopour, Hosein and Bayesteh, Alireza and Baligh, Hadi},
  year = 2014,
  month = sep,
  pages = {1--5},
  publisher = {IEEE},
  address = {Vancouver, BC, Canada},
  doi = {10.1109/VTCFall.2014.6966170},
  urldate = {2026-02-20},
  isbn = {978-1-4799-4449-1},
  langid = {english}
}

@article{wenDesigningEnhancedMultidimensional2022,
  title = {Designing {{Enhanced Multidimensional Constellations}} for {{Code-Domain NOMA}}},
  author = {Wen, Haifeng and Liu, Zilong and Luo, Qu and Shi, Chuang and Xiao, Pei},
  year = 2022,
  month = oct,
  journal = {IEEE Wireless Commun. Lett.},
  volume = {11},
  number = {10},
  pages = {2130--2134},
  issn = {2162-2337, 2162-2345},
  doi = {10.1109/LWC.2022.3194604},
  urldate = {2026-02-20},
  copyright = {https://ieeexplore.ieee.org/Xplorehelp/downloads/license-information/IEEE.html},
  langid = {english}
}

@article{yuDesignAnalysisSCMA2018,
  title = {Design and {{Analysis}} of {{SCMA Codebook Based}} on {{Star-QAM Signaling Constellations}}},
  author = {Yu, Lisu and Fan, Pingzhi and Cai, Donghong and Ma, Zheng},
  year = 2018,
  month = nov,
  journal = {IEEE Trans. Veh. Technol.},
  volume = {67},
  number = {11},
  pages = {10543--10553},
  issn = {0018-9545, 1939-9359},
  doi = {10.1109/TVT.2018.2865920},
  urldate = {2026-02-20},
  copyright = {https://ieeexplore.ieee.org/Xplorehelp/downloads/license-information/IEEE.html},
  langid = {english}
}

@article{yuSparseCodeMultiple2021,
  title = {Sparse {{Code Multiple Access}} for {{6G Wireless Communication Networks}}: {{Recent Advances}} and {{Future Directions}}},
  shorttitle = {Sparse {{Code Multiple Access}} for {{6G Wireless Communication Networks}}},
  author = {Yu, Lisu and Liu, Zilong and Wen, Miaowen and Cai, Donghong and Dang, Shuping and Wang, Yuhao and Xiao, Pei},
  year = 2021,
  month = jun,
  journal = {IEEE Comm. Stand. Mag.},
  volume = {5},
  number = {2},
  pages = {92--99},
  issn = {2471-2825, 2471-2833},
  doi = {10.1109/MCOMSTD.001.2000049},
  urldate = {2026-02-20},
  copyright = {https://ieeexplore.ieee.org/Xplorehelp/downloads/license-information/IEEE.html},
  langid = {english}
}

@inproceedings{sharmaSCMACodebookBased2018,
  title = {{{SCMA Codebook Based}} on {{Optimization}} of {{Mutual Information}} and {{Shaping Gain}}},
  booktitle = {2018 {{IEEE Globecom Workshops}} ({{GC Wkshps}})},
  author = {Sharma, Sanjeev and Deka, Kuntal and Bhatia, Vimal and Gupta, Anubha},
  year = 2018,
  month = dec,
  pages = {1--6},
  publisher = {IEEE},
  address = {Abu Dhabi, United Arab Emirates},
  doi = {10.1109/GLOCOMW.2018.8644383},
  urldate = {2026-02-21},
  isbn = {978-1-5386-4920-6},
  langid = {english}
}

@inproceedings{caiMultiDimensionalSCMACodebook2016,
  title = {Multi-{{Dimensional SCMA Codebook Design Based}} on {{Constellation Rotation}} and {{Interleaving}}},
  booktitle = {2016 {{IEEE}} 83rd {{Vehicular Technology Conference}} ({{VTC Spring}})},
  author = {Cai, Donghong and Fan, Pingzhi and Lei, Xianfu and Liu, Yingjie and Chen, Dageng},
  year = 2016,
  month = may,
  pages = {1--5},
  publisher = {IEEE},
  address = {Nanjing, China},
  doi = {10.1109/VTCSpring.2016.7504356},
  urldate = {2026-02-21},
  isbn = {978-1-5090-1698-3},
  langid = {english}
}

@misc{altera_scma_env1,
  author       = {{Altera Innovate Asia}},
  title        = {1st 5G Algorithm Innovation Competition -- Env1.0-SCMA},
  howpublished = {\url{http://www.innovateasia.com/5G/en/gp2.html}},
  note         = {Presentation, Online},
  year         = {n.d.}
}

\end{document}